\documentclass{elsart}

\usepackage{graphicx}

\newcommand{\mm}{\ensuremath{\mathrm{mm}}}
\newcommand{\mum}{\ensuremath{\mu \mathrm{m}}}
\newcommand{\cm}{\ensuremath{\mathrm{cm}}}
\newcommand{\cmsq}{\ensuremath{\mathrm{cm}^2}}
\newcommand{\scinot}[2]{\ensuremath{#1\!\times\!10^{#2}}}

\begin{document}

\begin{frontmatter}



\title{Precision planar drift chambers and cradle for
the TWIST muon decay spectrometer\\
}

\author[TRIUMF]{R.S. Henderson\corauthref{cor}},
\corauth[cor]{Corresponding author. Address: TRIUMF, 4004 Wesbrook
Mall, Vancouver, BC V6T 2A3, Canada. Tel: 604-222-1047. Fax:
604-222-1074. Email: rhend@triumf.ca.}
\author[TRIUMF,RRC]{Yu.I. Davydov},
\author[TRIUMF]{W. Faszer},
\author[VALPO]{D.D. Koetke},
\author[RRC]{L.V. Miasoedov},
\author[TRIUMF]{R. Openshaw},
\author[UOFA]{M.A. Quraan},
\author[UOFA]{J. Schaapman},
\author[RRC]{V. Selivanov},
\author[TRIUMF]{G. Sheffer},
\author[VALPO]{T.D.S. Stanislaus},
\author[RRC]{V. Torokhov}

\address[TRIUMF]{TRIUMF, Vancouver, BC, V6T 2A3, Canada}
\address[RRC]{RRC ``Kurchatov Institute'', Moscow, 123182, Russia}
\address[VALPO]{Valparaiso University, Valparaiso, IN 46383, USA}
\address[UOFA]{University of Alberta, Edmonton, AB, T6G 2J1, Canada}

\begin{abstract}

To measure the muon decay parameters with high accuracy, we
require an array of precision drift detector layers whose
relative position is known with very high accuracy. This article
describes the design, construction and performance of these
detectors in the TWIST (TRIUMF Weak Interaction Symmetry Test)
spectrometer.

\end{abstract}

\end{frontmatter}

\section{Introduction}
\label{Introduction}

Muon decay involves only the weak interaction and is a major
input to the Standard Model. The TWIST (TRIUMF
experiment E614) results will provide a test of these inputs and
provide an excellent window in which to search for physics beyond
the Standard Model. The aim of the TWIST experiment is to measure
the muon decay parameters $\rho$, $\delta$, $\xi$, and $\eta$
from muon decays with accuracies 10 to 40 times better than
existing data \cite{PDG}.  This will be achieved by measuring,
with high precision, the energy and angle distribution of
positrons (over a wide range) from the decay of polarized
muons. TWIST utilizes the M13 beam line at TRIUMF to transport
beams of 29.6 MeV/c surface muons from pion decay-at-rest into
the TWIST spectrometer. These polarized muons \cite{Pifer} pass
through a gas degrader and a foil degrader, which fine tune the
muon energy, so that they pass through the first half of the
spectrometer and stop in a target at the centre.

TWIST will measure, for the first time, all muon decay parameters
simultaneously, and will do so by recording two dimensional
(momentum and angle) distributions of the decay positrons.

\section{Overview}
\label{Overview}

Figure 1 is a conceptual view of the TWIST spectrometer. The
superconducting solenoid has an inner diameter of 100 cm and a
length of 223 cm. There are eight drift-chamber (DC) modules in
each half of the spectrometer, for a total of 44 drift
planes. These 16 DC modules are the main tracking elements of the
TWIST spectrometer. At the upstream and downstream ends of the
stack there are two proportional chamber (PC) modules, each
having four MWPC planes. The target module, at the centre of the
spectrometer, is a somewhat similar MWPC, but with the target
foil acting as the central cathode. The twenty-two drift chamber
layers and six MWPC layers are positioned both upstream and
downstream of the muon stopping target, in a highly symmetrical
pattern.

\begin{figure}
\begin{center}
\includegraphics*[width=120mm]{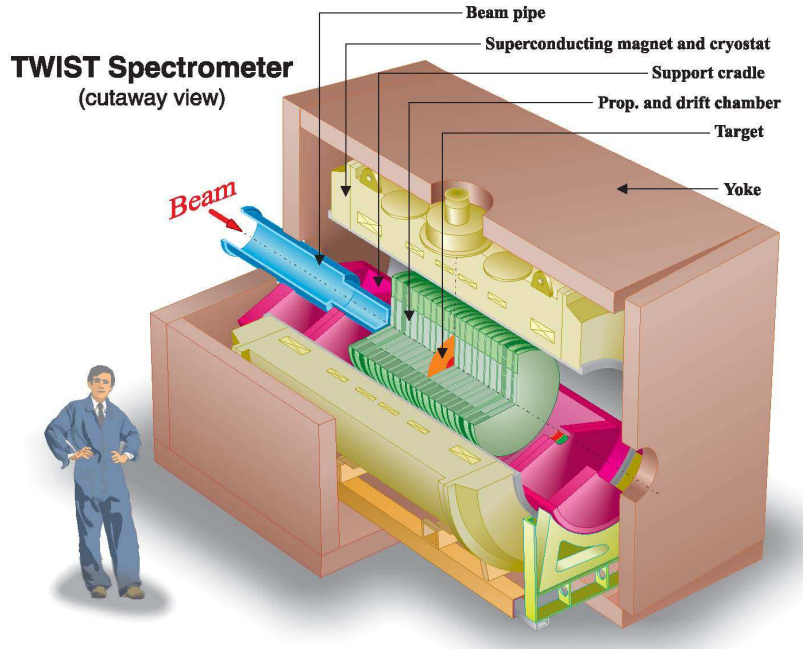}
\end{center}
\caption{
Conceptual drawing of the TWIST spectrometer. It 
shows the superconducting solenoid within the steel yoke, 
with the drift chambers and proportional chambers 
symmetrically placed from the central target.
}
\end{figure}

An external steel yoke was required to produce the highly uniform
two Tesla axial field for the DC tracking volume. This yoke was
modeled with OPERA-3d, then fabricated. It is 20 \cm\ thick at the
top and sides, and 8 \cm\ thick at the ends. The downstream end of
the yoke is hinged for easier access.

Figure 2 shows two field maps for the typical operating strength
of 2.00 Tesla. The upper plot shows the field map on the beam
axis ($x=y=0$), the lower plot shows the field at a radius of
165\ \mm, at the edge of the tracking volume. Within this
tracking DC volume ($|z|<500\ \mm$, $r<160\ \mm$), the
measurements determine the variations of the field as a function
of position to $\pm 1$ gauss. It is uniform over the full volume
to 80 G (full width). It has also been mapped at 1.96 Tesla and
2.04 Tesla.

\begin{figure}
\begin{center}
\includegraphics*[width=77mm]{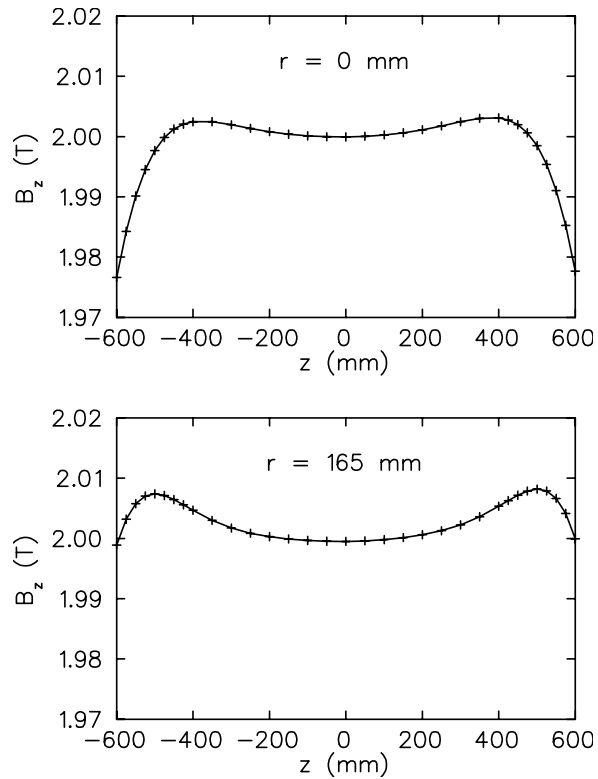}
\end{center}
\caption{
Maps of axial magnetic field, showing $B_z$ vs. $z$ 
for $x = y = 0$ (upper) and $r = 165\ \mm$ (lower).
}
\end{figure}

Figure 3 shows a side section view of the TWIST detector stack
and the cradle that contains them. The 19 modules are compressed
against the upstream end of the cradle by four custom built
pneumatic cylinders (see Sections 3 and 8).

\begin{figure}
\begin{center}
\includegraphics*[width=130mm]{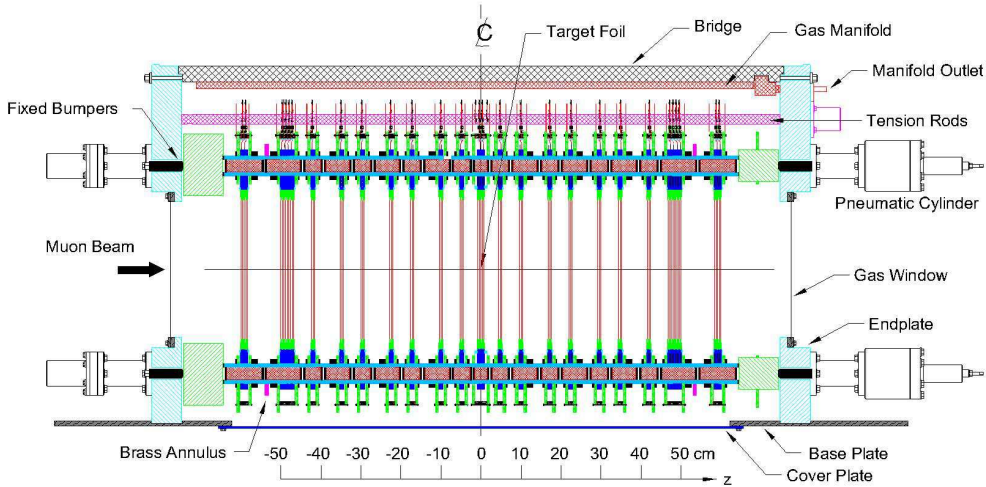}
\end{center}
\caption{
Side view of the TWIST cradle, showing thick FR4 
(flame resistant G10) pieces and stack of 19 modules, all 
pushed against the upstream cradle endplate by four 
pneumatic cylinders.
}
\end{figure}

To measure the muon decay parameters with the proposed
precisions, specific requirements must be met for; chamber
resolution, precision of individual chamber construction, and
precision of wire-plane positioning (especially in the $z$
direction).

With 44 drift-chamber layers distributed over a tracking length
of 1,000\ \mm, the positioning of the wire planes and thermal
effects were major concerns for the spectrometer design. Figure 4
shows a front section view of a DC module within the cradle and
magnet rail structure (only a few wires shown). A V layer is
shown within the circular FR4 (flame resistant G10) gas box, and the
coordinate system is indicated.

\begin{figure}
\begin{center}
\includegraphics*[width=110mm]{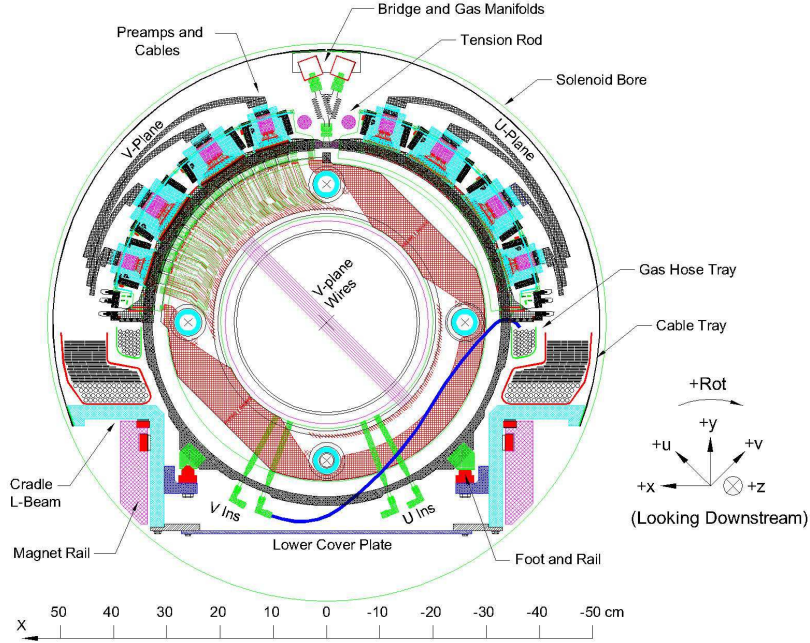}
\end{center}
\caption{
Front view of DC module in the cradle and 
magnet rail structure (looking downstream). The coordinate 
system is indicated.
}
\end{figure}

To position each wire plane and cathode foil as accurately as
possible in the $z$ direction, each module layer (including gas
box base and lid) contains a set of four high precision ceramic
spacers of a Russian material known as Sitall CO-115M
\cite{Sitall} (similar to Zerodur \cite{Schott}). These materials
were developed for telescopes and are made from a mixture of
glass in two different states. The resultant material has an
extremely small coefficient of thermal expansion
($\sim\!\scinot{1}{-7}\ (dL/L)/^{\circ}\mathrm{C}$) and can be
machined and polished to optical flatness. This, combined with
Sitall's good strength (elastic modulus $\sim\!\scinot{5.7}{10}$
Pa) allows us to accurately determine each wire plane's $z$
position.

Monte Carlo \cite{GEANT} estimates show that to measure the muon
decay parameters with an accuracy of about $\scinot{2}{-4}$ a
systematic error of not more than $\scinot{2}{-4}$ in
reconstructed muon decay positron energy, $E_e$, and angle,
$\cos\theta$, is also required. The energy losses of the muon
decay positrons along the $z$ axis of the spectrometer are
proportional to $1/\cos\theta$ and do not depend on the ($x$,$y$)
positions of the stopped muons. This means that the true maximum
positron energy, due to a muon decay, from a muon stopped in the
target $E_0^{\max} = 52.8$ MeV can be calculated from the
equation $E_{\mathrm{meas}}^{\max}(\theta) = E_0^{\max} - k/\cos\theta$
(where $k$ is an energy loss at $\theta =0$). Therefore an
absolute energy calibration of the spectrometer can be defined
from the muon decay spectrum itself. However, it is essential
that the detector system is very precisely
manufactured. Following is an estimate of systematic errors due
to manufacturing accuracy. The major source of problems would be
a systematic error in our length scales in the $u$, $v$, or $z$
directions.

The Monte Carlo study shows that a systematic shift in the
location of the U wires in the $z$ direction $\Delta U_z$ of
$\Delta U_z /U_z^{\mathrm{norm}} = |U_z -
U_z^{\mathrm{norm}}|/U_z^{\mathrm{norm}} = \scinot{5}{-4}$ causes
a bias in the three muon decay parameters of $|\rho - 0.75| =
\scinot{1}{-4}$, $|\delta - 0.75| = \scinot{1.1}{-4}$, and
$|{P}_{\mu}\xi - 1| = \scinot{6}{-5}$.  Since we determine the
positions of all our wire planes in $z$ to $<50\ \mu\mathrm{m}$
(see Section 3), our error $\Delta U_z / U_z^{\mathrm{norm}} <
50\ \mu\mathrm{m} / 1000\ \mm = \scinot{5}{-5}$ corresponds to a
muon decay parameter bias of $\ll 10^{-4}$. Sitall's very small
thermal coefficient makes TWIST effectively insensitive (in the
$z$ direction) to temperature. Sitall's good strength means that
the operating force of 1,470 N from a pneumatic cylinder
compresses the 1000\ \mm\ column of Sitalls (extent of DC modules)
by about $24\ \mu\mathrm{m}$ (see Section 8). The cylinders have
an uncertainty due to their O-ring friction of $\sim\!150$ N, but
this corresponds to only $\sim\!2.4\ \mu\mathrm{m}$, or $\Delta
U_z /U_z^{\mathrm{norm}} = 2.4\ \mu\mathrm{m} / 1000\ \mm =
\scinot{2.4}{-6}$.

A systematic shift of $\Delta U_u /U_u^{\mathrm{norm}}$ or
$\Delta U_v /U_v^{\mathrm{norm}}$ causes a similar bias in the
muon decay parameters. This corresponds to an uncertainty of the
pitch of the wires, or rather, the cumulative uncertainty in wire
position across the 320\ \mm\ active area of the wire planes. Our
wire planes are fabricated on glass substrates (see Section 6 for
fabrication details) and surveying of many wire planes indicates
that the width of the 320\ \mm\ wire plane varies only $\pm6\
\mu\mathrm{m}$ (see Section 3). This variation corresponds to
$\Delta U_u /U_u^{\mathrm{norm}} = \Delta U_v
/U_v^{\mathrm{norm}} = 6\ \mu\mathrm{m} / 320\ \mm\ =
\scinot{1.9}{-5}$ which also leads to an insignificant bias in
the muon decay parameters. Over 320\ \mm, the glass plate expands
$\sim\!1.6\ \mu\mathrm{m}/^{\circ}C$, so $\Delta U_u
/U_u^{\mathrm{norm}}$ and $\Delta U_v /U_v^{\mathrm{norm}}$ are
not highly sensitive to temperature.

Monte Carlo studies also showed that the wire positions within
each wire plane should be known to less than $20\ \mu\mathrm{m}$
RMS. While the positions of the wire planes in ($u$,$v$) can be
verified, with calibration runs using high energy beam pions, the
angles are too small to determine their $z$ positions. It is
clear that the geometry of the TWIST spectrometer exceeds the
mechanical tolerances required for this demanding experiment.

\section{Chamber design}
\label{Chamber Design}

The layout and the materials of the TWIST drift chambers were
chosen to minimize the effect of multiple scattering and energy
loss of both the incoming muons and the positrons leaving the
target from muon decay. Low mass was also a key requirement of
the chambers because the incoming surface muons have a range of
only $\sim\!140\ \mathrm{mg}/\mathrm{cm}^2$ (carbon
equivalent). Helium/nitrogen ($\sim\!97\!:\!03$) flows through the
cradle and between the modules, and the first and last cathode
foils in each module were also required to act as the module gas
windows.

After several GARFIELD \cite{GARFIELD} studies, and successful
prototyping, we chose a simple multi-wire design, where each DC
chamber consists of eighty $15\ \mu\mathrm{m}$ diameter sense
wires at a pitch of 4.0\ \mm, plus two guard wires on each
side. The cathode-to-cathode distance is also 4.0\ \mm. The
cathodes are $6.35\ \mu\mathrm{m}$ thick doubly aluminized Mylar
foil. To keep them as flat as possible, all cathode foils were
stretched to a high tension of $>200$ N/m and a high precision
gas system was designed, which maintains the differential
pressure between the module gases and helium/nitrogen to $\pm 7$
mTorr (see Section 9).

Dimethylether (DME) was chosen as the working gas for DCs. It has
some very attractive properties; low Lorentz angle, low atomic
number, good resolution and excellent quenching abilities. Due to
concerns of wire chamber aging and materials compatibility
\cite{GASENC}, a long duration aging study was carried out to test
all component materials in single-wire test chambers with DME.
Suitable materials were found and this damage study has been
published \cite{Openshaw}.

All 44 DCs are identical (fabricated as X layers), but become U
or V layers depending on how they are mounted inside the
modules. U wires are tilted +45 degrees from vertical, (looking
downstream), V wires at -45 degrees. There are seven (UV) modules
on either side of the target module and an eight detector-layer
module (VUVUUVUV), designated the dense-stack (DS), at each end
of the DC tracking volume. The DC layers start at $|z| = 44\ \mm$
and extend to $|z| = 500\ \mm$.

At the upstream and downstream ends of the stack there are two PC
modules with configuration (UVUV). These play a central role in
our pattern recognition process. These fast detectors
(CF$_4$/isobutane) provide discrimination amongst the various
tracks that pass through the detector (muon, decay positrons,
additional muons and/or decays, beam positrons, delta rays,
etc.).

The target module, at the centre of the spectrometer, has a
somewhat similar (UVUV) configuration, but with the target foil
acting as the central cathode. This module plays an important
role. The first two detector layers (TG\#1 and TG\#2) impose
fiducial constraints on the ($u,v$) coordinates of the muon
stopping location. These allow us to know that all decays are
fully contained within the tracking region. The third TG\#3
provides a veto so that any muons that stop past the target foil
are eliminated. The muon yields in TG\#1 vs TG\#4 (from the tails
of the stopping distribution) provide a sensitive estimate of the
mean stopping location in the target foil, to a precision of
$\sim\!10\ \mum$.

The target and PC modules have a sense-wire pitch of 2\
\mm. CF$_4$/isobutane was chosen for these modules,
since it has high drift velocity and Lorentz angle is not a
concern. The target layers extend from $z = -8$ to $z = +8\
\mm$. The PC layers start at $|z| = 584\ \mm$ and extend to $|z|
= 600\ \mm$.

Each wire plane is fabricated on a 3.18\ \mm\ thick 60 \cm\
diameter circular glass plate, having a thin printed circuit
board (PCB) laminated on the top surface. The four 4\ \mm\ thick
Sitalls are glued into holes in the glass plate (see Figs. 4 and
5). The wires are strung at the top Sitall surface and a cathode
foil sub-assembly positioned at the mid-Sitall point. An extra
cathode foil is required to complete each module. This cathode
only layer (CO), is also fabricated on a glass plate and uses the
same cathode foil sub-assembly.

\begin{figure}
\begin{center}
\includegraphics*[width=130mm]{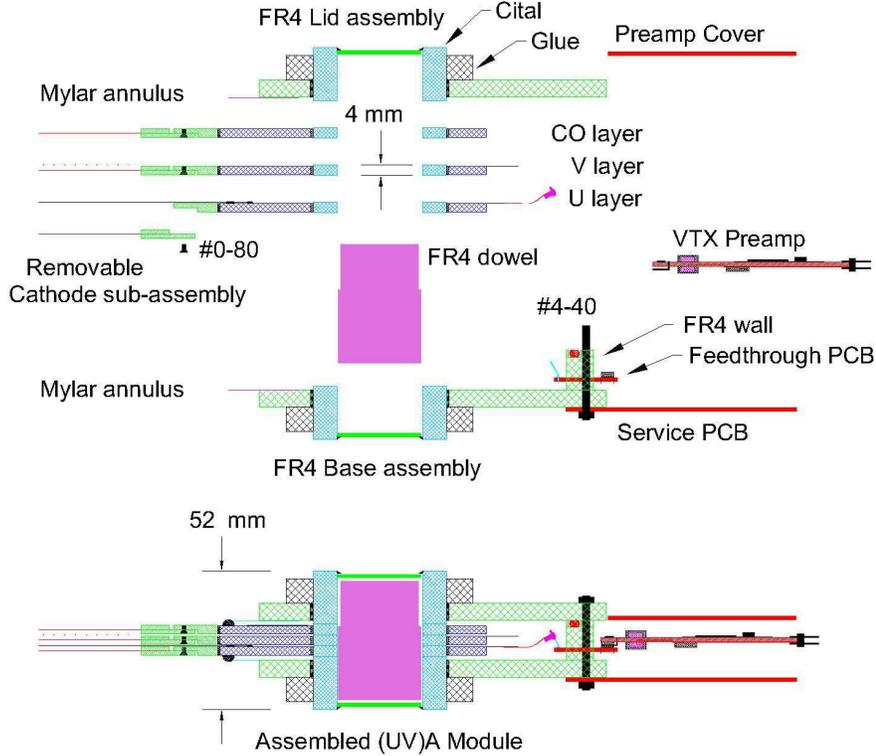}
\end{center}
\caption{
Section views of the (UV)A module (assembled 
lower and disassembled upper).
}
\end{figure}

As mentioned in Section 2, the gas box base and lid also contain
a set of four Sitall spacers. The result is that each module
layer is in contact at the Sitalls. In the cradle, the modules
touch each other only at their first and last Sitall
surface. When the stack of 19 modules is pressed towards the
upstream cradle endplate, the force ensures good Sitall-to-Sitall
contact (see Section 8).

Our Sitalls typically had surfaces which were flat and parallel to
$<0.5\ \mu\mathrm{m}$. Within the thickness groups (4, 8, 20 and
40\ \mm), the variation in individual Sitall thickness was
$\sim\!3\ \mu$m. All Sitalls thicknesses were measured and they
were arranged in sets of four. Within each set the thickness
variation was $<0.5\ \mum$. The Sitall thickness data is used to
determine the position of each wire plane. In addition, the
compression of the Sitalls (in the cradle) frictionally locks each wire plane
position in the ($u$,$v$) coordinate plane. The actual position
of each wire plane is determined by fitting straight tracks from
120 MeV/c pions during calibration runs (see Section 12).

With the wire planes fabricated on glass substrates, the
($u$,$v$) positions of wires is less sensitive to temperature (as
mentioned in Section 2). With a coefficient of thermal expansion
of $\sim\!\scinot{5}{-6}\ (dL/L)/^{\circ}C$, the 320\ \mm\ wide
wire planes expand $\sim\!1.6\ \mum/^{\circ}C$.

Figure 4 shows a V layer within the circular FR4 gas box. The 80
sense wire signals pass through the gas box wall to nearby
preamps, the output signals travel on mini-coax cables collecting
in a cable tray on either side. These cable trays are removable,
to allow removal of the cradle and detectors from the seven racks
of readout electronics and other services.

The many 0.25'' polyethylene gas input lines (one per detector layer)
collect in two smaller permanent trays. These gas lines pass
between and under the modules to inlet fittings that are
accessible when the lower cover plate is removed. Within the
modules, flat polyester ``straws'' inject the gas between the
detector layers to the active areas. The module gas outlets are
at the top and connect with soft neoprene bellows to two gas
manifolds, one for DME, the other for CF$_4$/isobutane.

Each module has two 6.35\ \mm\ thick FR4 feet, one flat and one
V-shaped. These feet are doweled and screwed to the gas box base,
they support and position the modules in the spectrometer by
means of flat and V-shaped rails within the cradle support
structure (see Section 8). Two large longitudinal L-shaped beams
give the cradle great stiffness. The cradle rolls in and out of
the solenoid on the large magnet rails (see Section 8).

During data taking, the 19 modules are immersed in
helium/nitrogen ($\sim\!97\!:\!03$) gas. The addition of the small
amount of nitrogen is necessary to prevent HV breakdown on the
module exteriors. Each module was tested for HV breakdown in this
gas mixture as part of the fabrication quality control (QC).

\section{The (UV)A module}
\label{UVA}

Figure 5 shows two section views of the (UV)A 
module (one disassembled). Section 5 describes 
the four other module types (two DC and two 
PC), but many of the design features of the 
(UV)A are common to all the module types. 

The (UV)A module consists of five component 
layers:
\begin{enumerate}
\item FR4 gas box base.
\item U plane (U).
\item V plane (V).
\item Cathode-Only plane (CO).
\item FR4 gas box lid.
\end{enumerate}


\subsection{Gas box base}

The gas box base consists of a circular FR4 plate (6.35\ \mm\
thick and $\sim\!70\ \cm$ diameter). This plate has curved FR4
wall pieces laminated to it, building a full circular wall. In
the U and V readout arcs, the wall contains 1.6\ \mm\ thick
feedthrough printed circuit boards (PCBs), which transfer the 80
sense-wire signals (at HV) from the detector layer to the two
24-channel and two 16-channel preamplifiers mounted just outside
the wall. These feedthrough PCBs also bring HV and pulser signals
into the module.

A set of four 20\ \mm\ long Sitalls is also glued into holes in the
base plate. Since the base and lid are bolted together at the
O-ring seal, we were concerned that warps in the 6.35\ \mm\ thick FR4
plates would result in the gas box warping on final assembly. To
avoid this, bases and lids of gas boxes are not swapped around,
but stay as a pair. They are bolted together first, then the
Sitalls are glued into the base. Finally, stacks of Sitalls are
added inside the gas box to give the correct internal distance
(12\ \mm\ for the (UV) modules). The Sitalls are then glued into the
lid. In this way, the assembled gas box should have coplanar
Sitall surfaces. Approximately 1\ \mm\ inside the
outer surface of these Sitalls, 1.6\ \mm\ thick FR4 disks are
permanently glued, making a gas seal. The top surface of the gas
box wall contains an O-ring in a groove to seal against the gas
box lid.

An FR4 dowel is inserted into each base Sitall, forming four
posts onto which the other component layers are installed. For
the (UV)A module, these four dowels protrude 11\ \mm\ into the
module, then narrow and fit more loosely into the lid plate
Sitalls.

The Sitall IDs are $32.05 \pm 0.05\ \mm$, and the dowel ODs are
$31.93 \pm 0.01\ \mm$. If each of the four Sitalls in each
detector layer were positioned perfectly, this would allow a
relative movement of base, U and V layers of $120 \pm 60\
\mum$. In practice the movement was $<80\ \mum$.

The gas box wall also contains two 0.25'' Swagelok gas inlet
fittings (one for each wire plane) and a 0.5'' Swagelok fitting
for the module gas outlet. Two 1.6\ \mm\ thick service PCBs are
bolted to the gas box base. Each covers $\sim$82 degrees of arc
in the area of the preamps and extend $\sim\!60\ \mm$ beyond the
gas box wall. These PCBs have preamp mounting guides, grounding
and the preamp power bus (+4\ V).

It would have been convenient if the gas box base (and lid) could
have had their own window foils glued across their central
cutouts. Unfortunately, this option was ruled out by
considerations of multiple scattering and the mass budget for the
incoming muons. Instead, a $100\ \mum$ thick Mylar annulus is
permanently glued to the inner edge of these two FR4
plates. These Mylar pieces protrude 40\ \mm\ into the central
cutout of the FR4 plates, but stop 4\ \mm\ outside the active
area. After the module is assembled, these two Mylar pieces are
glued to the U layer glass plate and CO layer PCB with latex
rubber. This forms a mechanically soft gas seal and is easily
removable. Latex rubber was chosen, after tests with other
synthetic rubbers (silicones and urethanes) showed extremely high
aging rates with DME ($>1000 \%/\mathrm{C}$ per cm of wire)
\cite{Openshaw}.

\subsection{U plane}

The U drift plane, as shown in Figs. 4 and 5, consists of one
wire plane and one cathode foil sub-assembly.

This layer has four 4\ \mm\ thick Sitalls positioned accurately
and glued into holes in a 3.18\ \mm\ thick 60 \cm\ diameter
circular glass plate. On the upper surface of the glass plate, a
thin (180\ \mum) PCB is laminated. This thin PCB has the
solder-pad pattern for the wires, and traces from the pads to
where flexible output Kapton/Cu ribbon cables are soldered. When
the U drift plane is positioned inside the gas box base (on the
FR4 dowels) the ribbon cables are plugged into connectors on the
gas box feedthrough PCB. No solder connections are
required. Section 6 describes the fabrication of the wire plane
in detail.

There is an attached cathode foil sub-assembly within the central
cutout of the glass plate (see Section 6). Both surfaces of the
aluminized Mylar foil are clamped against copper surfaces and
cross-connected. This foil connection is also brought out on two
Kapton/Cu ribbon cables to the gas box wall PCB.

\subsection{V plane}

The V drift plane is identical to the U drift plane, simply
positioned at 90 degrees to the U plane and on top of it.

\subsection{Cathode-only plane}

The cathode-only plane (CO) is similar to wire plane layers, in
that it has four 4\ \mm\ Sitalls, a glass plate and a cathode foil
sub-assembly. The PCB is however far simpler, since there is no
wire plane and only two Kapton/Cu ribbon cable connections of the
foil to the gas box wall PCB.

With only $\sim\!0.5\ \mm$ between the glass/PCB layers of the U,
V and CO sub-layers, proper flushing of the two approximately 103
cc active volumes was a significant design concern. Our solution
was to make flat polyester-film straws, that directly inject the
chamber gas into these active volumes. These straws have a heat
sealed edge and are permanently glued to the inlet gas fitting on
the inside surface of the gas box wall. They are 10\ \mm\ wide in
the flat region and extend to the active volumes of
the U and V plane. During module assembly, the two straws are
easily bent out of the way, the U plane installed, its straw laid
down on it, then the V plane and its straw, then the CO
plane. This simple and reliable system ensures good flushing of
the active areas of the chambers.

\subsection{Gas box lid}

The gas box lid plate is simpler than the gas box base, with only
four 20\ \mm\ long Sitalls and the 6.35\ \mm\ thick FR4 sheet
(with Mylar annulus). As mentioned in Section 4.1, these Sitalls
are glued in the FR4 lid plate in association with the gas box
base.  The lid is bolted down to the gas box base, forming a gas
seal against the O-ring in its groove. As with the gas box base,
there are 1.6\ \mm\ thick FR4 discs glued into the ends of the
Sitalls to form gas seals. After module assembly, a latex rubber
seal is made to the nearest detector layer (the CO).

\section{The other four module types}
\label{othermodules}

The (UV)A drift module was described in Section 4. There are four
other module types in the TWIST spectrometer, they are all based
on the same simple structure. They are:

\subsection{(UV)B module}

This is also a drift chamber module (DC) using DME gas. It is
almost identical to the (UV)A module. The only difference is that
it has 40\ \mm\ long Sitalls in the gas box base instead of 20\
\mm. By alternating (UV)A and (UV)B modules, for the first seven
modules on either side of the target, we produce a better
tracking chamber pattern.

\subsection{Dense stack}

This is also a drift chamber module using DME gas. Like the
(UV)B, it has 40\ \mm\ Sitalls in the gas box base and 20\ \mm\
Sitalls in the gas box lid. However, the wall of the gas box base
is approximately three times as tall and, in the two readout
regions, has feedthrough PCBs for four wire planes instead of
one. Instead of having three detector layers inside the gas box
(U,V and CO), the dense stack has nine detector layers (VUVUUVUV
and CO). The pattern break in the center is meant to reduce
tracking ambiguities. Of course, the FR4 dowels are 24\ \mm\
longer as well and there are eight gas inlet fittings and straws
instead of two.

\subsection{PC module}

This module uses four PC wire planes instead of two DC planes. It
has a taller gas box wall, with two feedthrough PCBs in each
readout area. There are four gas inlets and straws. The five
sub-layers in the gas box are UVUV and CO. The Sitalls in the gas
box lid and base are all 40\ \mm\ long.

The PC wire plane has 160 sense wires, and three guard wires each
side, all strung at 2\ \mm\ pitch. This gives an active area of
320\ \mm\ diameter. The central 32 wires are individually read
out to better handle the incoming muon beam and the remainder are
read out in groups of four wires, to reduce the number of readout
channels needed. There are 64 readout channels for each PC plane,
using four 16-channel preamps. CF$_4$/isobutane was chosen for
the PCs, since it has high drift velocity and the Lorentz angle
is not a concern.

\subsection{Target module}

The target module is the most specialized module in the TWIST
spectrometer. The five sub-layers in the gas box are UVUV and
CO. Each of the wire planes has its own cathode foil
sub-assembly, so there are five cathodes and four wire
planes. The major difference of the target module is that the
central cathode is also the experimental muon stopping target.

Each target wire plane has 48 sense wires and three guard wires
on each side, all strung at 2\ \mm\ pitch, and use
CF$_4$/isobutane gas. The 48 sense wires are all individually
read out, giving an active area of 96\ \mm\ diameter.

This module's main function is to define beam particles entering
and exiting the target. Since the axial field confines the low
emittance beam particles to small radii, the smaller active area
is sufficient.

The major target of interest for TWIST is high purity aluminum of
thickness about 70\ \mum. However, high purity aluminum
($>99.999\%$) foil was not available in the required size. Also,
aluminum is far less elastic than Mylar and tests indicated that
a foil 340\ \mm\ in diameter was not flat enough to act as a
wire-plane cathode. A smaller diameter sub-assembly was also not
an option, since we wished to keep the mass in this region as low
as possible. Therefore, for the initial study of the TWIST
spectrometer, the target was 125\ \mum\ thick Mylar with carbon
\cite{GC} on each side (thickness between 5 and 20\ \mum), for a
total of $\sim\!145\ \mum$, which has similar stopping power to
70\ \mum\ aluminum.

The target foil sub-assembly is fabricated similar to the usual
6.35\ \mum\ thick aluminized Mylar. The carbon-painted Mylar was
stretched and glued between the same two 1.6\ \mm\ thick FR4
discs to make a foil sub-assembly. When this sub-assembly is
attached to the second U plane, the foil is against the usual FR4
retaining ring, glued into the glass plate (see Section
6). Because of its greater thickness, this $\sim\!145\ \mum$ foil
is not perfectly centered between the adjacent V and U wire
planes. For the second U plane, the distance is the usual 1997\
\mum\ (2\ \mm\ minus half the foil thickness) and the wire plane is
symmetrically placed between the adjacent cathode foils. However,
for the first V plane, the wire plane to target foil distance is
$\sim\!139\ \mum$ less than usual and the wire plane is
effectively off-center. Since these wire planes are not used for
precision tracking, this was acceptable.

While the Mylar target was in use, we developed a new technique
for the aluminum target foil. This uses a stretched Mylar foil
with a central cutout. A small diameter aluminum foil is then
glued over this hole. The Mylar is 25\ \mum\ thick doubly
aluminized, with a 120\ \mm\ diameter hole, and effectively acts
as a spring to keep the aluminum foil tensioned and flat. The
aluminum foil is 150\ \mm\ in diameter, so there is a 15\ \mm\
overlap glue region. TRA-CON conductive silver epoxy is used to
ensure electrical connection to the Mylar foil. The aluminum foil
is $71 \pm 1\ \mum$ thick and the purity is $>99.999\%$
\cite{Goodfellow}. To avoid any problems with this glue region,
we also inserted a pair of 25\ \mum\ thick Kapton masks in each
of the wire planes adjacent the target foil. These masks
eliminate gas gain elsewhere and have a central cutout 110\ \mm\
in diameter. Only the central region of the target planes is of
interest (their active width is only 96\ \mm), so these masks do
not affect the target module performance.

The end result is a target module with a well-tensioned high
purity central aluminum target foil and an active area of about
100\ \mm\ diameter; but still having low mass out to 330\ \mm\
diameter. The resultant combination foil has worked very well. By
choosing which side of the Mylar the aluminum foil is glued on,
and using 39\ \mum\ thick spacers, we were able to position the
center of the aluminum foil mid-way between the V and U wire
planes. Now both these wire planes are closer to the target foil
than their other foil, but the difference is only 32\ \mum\ instead
of the 139\ \mum\ for the earlier Mylar/carbon target.

\subsection{Mirror modules}

The aim of the TWIST spectrometer is to be as symmetric as
possible with respect to the central target foil. We therefore
decided that the nine modules on either side of the target module
should be mechanically mirrored, i.e., they should all have their
gas box lid plates facing the central target.

With this in mind, the gas boxes were designed with removable
feet. These feet, one flat and one V-grooved, position the module
in the cradle. By switching the feet, and rotating about the
vertical axis, a gas box can be used in either position. This
effectively reduces the different module types from nine to
five. However, rotating a module like this effectively turns a U
plane into a V plane, and vice versa, introducing an
asymmetry. Therefore, while the components are the same, a module
is assembled differently if it is to be used in the upstream or
downstream half of the spectrometer. For example, a downstream UV
module has the first wire plane installed in the module gas box
tilted +45 degrees (looking into the open gas box). So a
downstream (UV) is assembled (Base, +45, -45, CO, Lid), while an
upstream (UV) is assembled (Base, -45, +45, CO, Lid).

\subsection{Spare modules}

We also produced five fully instrumented spare modules, one of
each type; (UV)A, (UV)B, dense stack, target and PC. In the two
years since the TWIST spectrometer was commissioned, the
performance of the TWIST modules has been excellent. All 3,520 DC
wires operated at full efficiency with no dead or hot wires. One
PC wire broke at the end of a long period of data taking, thereby
requiring the module to be removed and replaced with the spare PC
module. The damaged PC module was repaired and is now the
spare. The second target module allowed the aluminum target to be
installed and the module to be fully tested, ready for a quick
replacement in the spectrometer.

\section{Wire plane fabrication}
\label{planefab}

As mentioned in Section 3, a TWIST wire plane layer, consists of
one wire plane and one removable cathode foil sub-assembly. The
wire plane design is based on a 3.18\ \mm\ thick circular glass
plate. This plate has a diameter of $\sim\!600\ \mm$, with a
central cutout of $\sim\!398\ \mm$ diameter and four smaller
$\sim\!54\ \mm$ diameter holes positioned every 90 degrees and
260\ \mm\ from the center.

The steps to fabricating a wire plane are as follows:
\begin{enumerate}
\item Glue thin PCB (180\ \mum) on a glass plate.
\item Glue set of four 4\ \mm\ Sitalls into glass plate.
\item Glue FR4 cathode retaining ring into central
cutout of glass plate.
\item String wires above PCB surface.
\item Glue wires to glass plate (glue bumps), rotate
winding table to vertical, allow to set overnight.
\item Rotate winding table to horizontal, solder
wires to PCB. Trim and clean.
\item Measure wire tensions, replace if $\mathrm{T} < 26\
      \mathrm{g}$.
\item Put wire plane on milling machine. Mill glue
and solder bumps to $\leq 650\ \mum$ from glass.
\item Measure wire positions in $x$ and $z$.
\item Replace wires out of position $>20\ \mum$ in $x$.
\item Solder 19 flexible Kapton/Cu ribbon cables 
on readout edge of PCB.
\item Clean and store until module assembly.
\item Install cathode foil sub-assembly with \#0-80 
nylon screws.
\end{enumerate}

The removable cathode foil sub-assembly is fabricated by
stretching the 6.3\ \mum\ foil, then gluing it to a 1.6\ \mm\ thick FR4
annulus (ID=339\ \mm, OD=379\ \mm). When the glue sets, two small
dowel holes are cleaned and the Mylar pierced, then a second
1.6\ \mm\ thick FR4 annulus (ID=339\ \mm, OD=360\ \mm) is doweled and
glued to the other side of the Mylar. When trimmed and cleaned,
the resultant assembly is reasonably flat. The foil sub-assembly
is attached to the cathode retaining ring by 24 \#0-80 nylon
screws. The difference in ODs means that the foil surface is
clamped against the retaining ring, making the foil $z$ position
insensitive to the thickness of the FR4. Two ``notches'' in the
larger FR4 annulus allow both sides of the aluminized Mylar to be
clamped against copper surfaces, ensuring reliable low impedance
connections to the thin PCB laminated on the glass plate. Two
flexible Kapton/Cu cables bring these ground connections to the
gas box wall feedthrough PCB.

Although the wire planes are mounted as U or V planes in the
modules, during fabrication, they are considered as X planes
(wires vertical). For step (2), the set of four Sitalls are
clamped on a thick glass assembly table which is flat to $\pm
0.5\ \mum$. For step (3), the incomplete unit is again clamped on
the assembly table (only at the Sitalls), the FR4 retaining ring
is positioned on glass spacers in the cutout and glued in
place. While the wires are strung at the surface height of the 4\
\mm\ Sitalls, the cathode foil retaining ring is positioned as
accurately as possible 2\ \mm\ away, at the mid-thickness.

During stringing, the detector layer is clamped down (only at the
Sitalls) to a similar thick flat glass winding table. The wire
plane is strung using precision glass combs. Stringing was
carried out manually, in a class 1000 clean room. The room
temperature was held stable within $\pm 1^{\circ}C$. When
positioned from comb-to-comb, the wire is above the PCB surface,
so the height of the wires ($z$ position) is set by the glass
combs and is therefore precise relative to the Sitall surfaces.

All wire chambers use 15\ \mum\ diameter gold-plated
tungsten/rhenium W(Re) sense wires. Lengths of wires vary from 40
\cm\ in the center of a plane to 23 \cm\ on the edge. As each
wire is strung, one end is soldered to an external pad, then it
passes across the two glass combs, and is then tensioned and
soldered to another external pad. When the full plane of wires is
strung in this way, each wire is glued to the glass plate (just
beyond the cutout) with small epoxy glue beads. When this is
completed for the whole plane, the winding table is rotated to
the vertical position and the glue is allowed to set
overnight. In this way, the wire plane is strung horizontally,
but finally moved to the vertical, so that gravitational sags of
the glass plate do not affect wire positions in $z$. This matches
how the wire planes are positioned in the TWIST experiment. The
next day, the winding table is rotated back to the horizontal,
the wires soldered to the PCB pads and trimmed.

With only $\sim\!0.5\ \mm$ between the glass/PCB pieces of the detector
layers, the height of the wire solder bumps and glue beads was a
problem. Attempts to keep them shallow were a failure and several
glass plates were broken in early prototype modules. Our solution
was to place the wound wire plane in a milling machine (mounted
at the Sitalls), and machine any solder and glue surfaces that
were too high. In this way we guaranteed a clearance of $\geq 0.15\ \mm $.

All drift chamber layers (DCs) are the same, having 80 sense
wires, plus 2 guard wires on each side, with a pitch of
4.0\ \mm. The PC wire plane, has 160 sense wires, and an active
area of 320\ \mm\ diameter. The target wire plane has 48 sense
wires, and an active area of 96\ \mm\ diameter. The PC and target
planes both have 3 guard wires on each side and a wire pitch of
2.0\ \mm. All three wire plane types are wound on the same winding
table and with the same procedure.

The fully instrumented spectrometer has 19 modules and required
44 DC planes, 4 target planes, 8 PC planes and 19 CO planes. The
five spare modules contain another 12 DC planes, 4 target planes,
4 PC planes and 5 CO planes.

Quality control was carried out during all steps of
fabrication. This was essential to ensure reproducible results
during the production. The mechanical parameters of every
detector layer (dimensions of components, wire tensions and
positions, etc.) were measured during production and stored in a
data base.

For electrostatic stability at 2000\ V, only a 4\ g tension is
theoretically required for 40 \cm\ long 15\ \mum\ diameter tungsten
wires. We used 35\ g weights, well below the $\sim\!50$\ g typical
breaking tension of 15\ \mum\ W(Re).

Average measured wire tensions are about 31\ g, the difference
being due to the friction of wires on the winding equipment. The
distribution of measured tensions for all 6,784 sense wires on all
56 DC, 12 PC and 8 target layers had an RMS of only 1.25\ g. All
wires with tension less than 26\ g were replaced.

After stringing, machining of the solder and glue bumps, and
tension measurement, the position of each wire in the plane is
mapped. This is usually only done in the plane of the wires,
i.e., the $x$ value for two well separated values of $y$ (along the
wire). However, since we were going to so much effort to control
the $z$ positions, with glass surfaces and glass combs, etc., we
also wanted to measure the $z$ position of each wire (at two values
of $y$).

This mapping in $x$ and $z$ was achieved using a digital readout carriage 
with two CCD cameras. One camera observed the wire from directly
above the plane. The carriage was moved until the wire was
centered on a monitor with cross-hairs and the digital scale
readout gave $x$ directly. To measure the $z$ coordinate, a second
camera was mounted to view the wire from a 45 degree angle. Using
a second monitor and cross-hairs, the difference in the two
values of the digital readout gave the $z$ position. So, the ($x$,$z$)
wire positions on each wire were measured at two values of $y$,
20 \cm\ apart.

Figure 6 shows wire $x$ position residuals (distance from correct
location) for a typical DC wire plane, plus the distribution of
those $x$ residuals. Two measurements were made of this wire
plane, by different operators, eight months apart. The first
measurement was done immediately after winding. The second was
made after replacement of wire number 18 (shown as a star). This
figure shows the excellent reproducibility of the measured wire
positions and also of the wire replacement ($<3\ \mum$). Tests on a
glass ruler with precise diamond cut grooves indicate the
measurement accuracy is $\sim\!2.3\ \mum$ for $x$.

\begin{figure}
\begin{center}
\includegraphics*[angle=90,width=130mm]{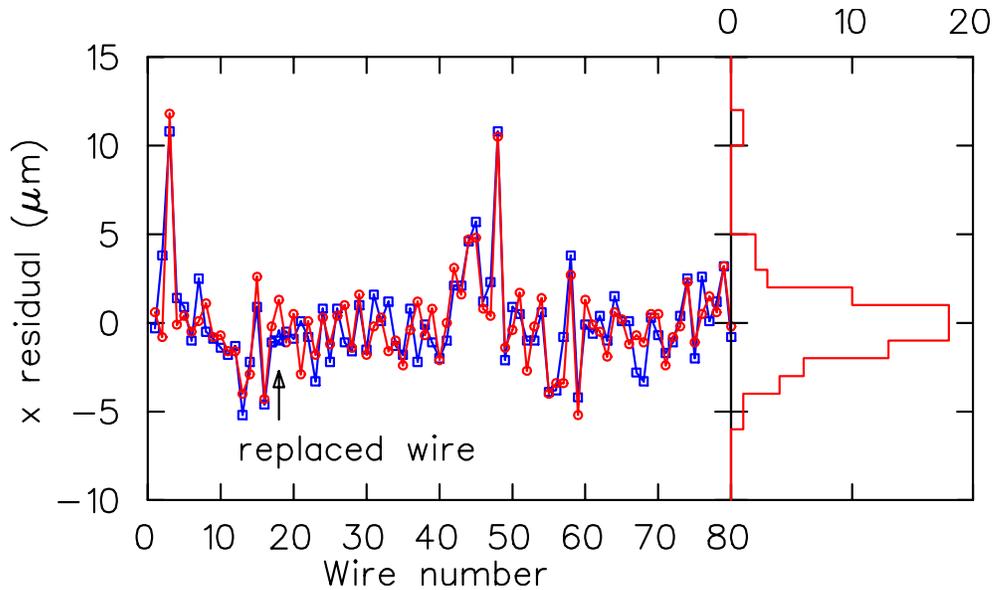}
\end{center}
\caption{
Wire $x$ position residuals (distance from correct location) for
a typical DC wire plane. Blue squares are measurements soon after
fabrication, red circles are data remeasured 8 months later.  }
\end{figure}

Figure 7 shows the distribution of measured $x$ wire position
residuals summed for 6,304 wires (from 50 DC planes, 12 PC and 8
target layers), with $\sigma =3.3\ \mum$. 
These residuals were from the readout side of the wire plane;
similar results were obtained on the nonreadout side (20\ \cm\ away
in $y$). A total of 77 wire planes were fabricated (including those
in the spare modules), having 6,944 wires. Our database contains
all measured ($x,z$) values. 

\begin{figure}
\begin{center}
\includegraphics*[width=77mm]{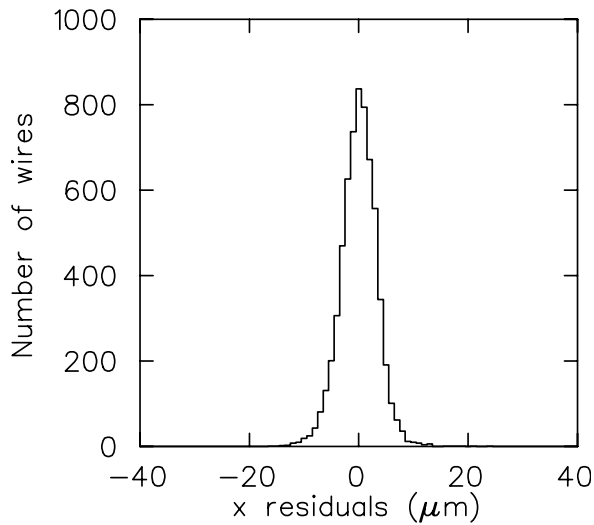}
\end{center}
\caption{
Wire $x$ position residuals summed for 6,304 sense wires from the
readout side of 70 wire planes (50 DC, 12 PCs and 8 target
planes), with $\sigma = 3.3\ \mum$.
}
\end{figure}

For the 77 wire planes, there are very few wires out of position
more than 15\ \mum\ in $x$. As part of our QC, we replaced any wire
more than 20\ \mum\ out of position. Only three were missed, all
less than $25\ \mum$. The results demonstrate that
our winding equipment has provided high quality wire plane
production, despite using many people over a two year production
period.

Similarly, Fig. 8 shows the wire position residuals for a typical
PC plane in the $z$ direction (negative being down). The upper
part shows the results for the plane unclamped, the lower part
the same wire plane clamped. By clamped, we mean there was force
applied to each Sitall, pushing it down against the granite
surface of the scanning table. The change is significant. One can
see that while the wire-to-wire variation is similar to the $x$
measurement, the range of the $z$ residuals is much larger,
$\sim\!60\ \mum$ (for unclamped) compared to $\sim\!19\ \mum$ for
the $x$ residuals of Fig. 6.

\begin{figure}
\begin{center}
\includegraphics*[width=77mm]{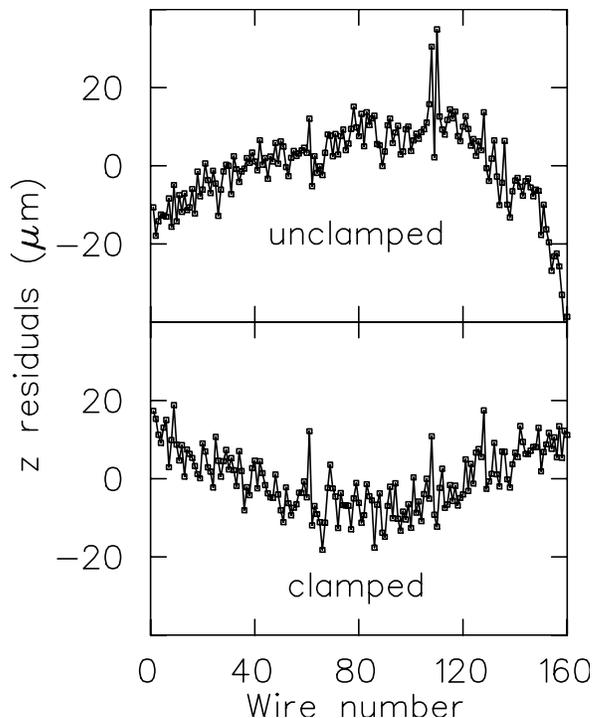}
\end{center}
\caption{
Wire position $z$ residuals for PC wire plane (\#153) 
without clamping (upper) and with clamping (lower).
}
\end{figure}

This is understood as follows. During the final stage of
stringing (when the glue beads are setting) the wire plane was
vertical and bolted firmly (at the Sitalls) to a solid and flat
glass plate. However, during the ($x$,$z$) wire position mapping,
the wire plane was horizontal and not clamped down. It was
therefore subject to flexing and gravitational sag. The
calculated gravitational sag of the glass plates (midway between
Sitalls) is 17\ \mum. But we see in the upper part of Fig. 8,
that this unclamped wire plane is not sagging down, but flexing
up. The maximum bending of the glass plates due to wire tension
loads was calculated using ANSYS \cite{ANSYS} to be only 1.3\
\mum\ for a DC plane and 2.6\ \mum\ for a PC plane, this is not
significant.

It was found that, because the glass plates were slightly warped,
there had been concern that they would touch each other, since
the gaps between detector layers are only $\sim\!0.5\ \mm$. To avoid this,
weights had been placed on the glass plates during their gluing
to their set of Sitalls (step (2) in list). Naturally, when
released, they sprang back to their natural shape. During gluing
of the cathode retaining rings and stringing, the planes were
again clamped to the assembly and winding tables respectively,
but during ($x$,$z$) measurement they were unfortunately not clamped.

Figure 8 demonstrates that, when the same wire plane was clamped,
the glass plate was forced much flatter again (lower part of
figure). It now shows the center of the wire plane $\sim\!18\ \mum$ lower
than the edges, in good agreement with the calculated
gravitational sag of 17\ \mum. Unfortunately, this wire plane was
the only one not in the experiment or in spare modules, so other
clamped $z$ residual data could not be obtained.

Tests indicate that a load of only 50\ N is required on each Sitall
to clamp the glass plate to its flattened state. In the cradle,
the Sitall columns are compressed with 1,470\ N, so we believe all
the wire planes are certainly flattened. Also, since the planes
are vertical, the 17\ \mum\ gravitational sag should not be
present. Our existing $z$ maps allow us to determine if any wire is
out of $z$ position by more than $\sim\!10\ \mum$ (relative to its
neighbors). As part of our QC we replaced any wire that was out
of $z$ position by more than $\sim\!25\ \mum$ (relative to its
neighbors). Only four wires out of position by more than $50\
\mum$ in $z$ were missed.

Few of our 77 wire planes are badly warped in $z$. The wire plane
shown in Fig. 8 is among the worst. Most have warps $<20\ \mum$. The
summed data of $z$ residuals has $\sigma = 8.4\ \mum$. This is
quite low, considering that even a clamped wire plane has a
gravitational sag of 17\ \mum\ (calculated).

\section{Bench tests}
\label{BenchTests}

All modules were bench tested before use in the TWIST
spectrometer. Gas gain was measured across each wire plane using
an argon/isobutane (25:75) gas mixture and $^{55}\mathrm{Fe}$
X-rays. This mixture has an operating voltage similar to that of
DME.

These tests showed that signal pulse height uniformity is within 20\%. The
variation is mostly due to increase of anode-cathode gaps in the
centers of planes because the outer cathode foils also serve as
gas windows and differential pressure between chamber gas and
environment was not perfectly zeroed for these tests. Another
test was carried out with a 10 mCi uncollimated
$^{90}\mathrm{Sr}$ source in order to check the ability of each
plane to hold high voltage with current up to 50 $\mu$A for at least
30 seconds.

Modules were also tested for gas tightness with helium. For (UV),
target, and PC modules, a leak rate less than $<0.3$ cc/min was
considered acceptable. For the dense stack modules, leak rates
$<0.6$ cc/min were accepted (see Section 9.3 for details).

Modules were also tested for HV stability in helium/nitrogen
($97\!:\!03$). This was done with pure isobutane in the module to
over-quench them and allow testing at higher than operating
voltage. All assembled modules have passed these bench tests.

\section{Module support - the cradle}
\label{cradle}

The TWIST solenoid has a bore diameter of 1050\ \mm. To position
the nineteen TWIST detector modules, the cradle was designed to
be as rigid as possible, so the positions of wires (in $x,y,z$) are
well understood and constant during long periods of data
taking. As was seen in Fig. 4 (the front view of DC module in
cradle structure), the L-shaped cradle beams and the magnet rails
were made as large as possible while allowing the entire
structure to be moved up to 10\ \mm\ (in any direction) within the
solenoid bore. The cradle and magnet rails are made of aluminum
(6061 T6). All screws, bolts and dowels are of non-magnetic
titanium or brass.

Figure 9 shows a 3d drawing of the cradle and magnet rail structure
(not true colors). The two magnet rails are $5\ \cm \times 18\ \cm$
and 265 \cm\ long. The downstream ends of the rails come almost to
the magnet door. When the magnet door is open, an external cart
having similar profile beams can be connected to the magnet rails
to allow the cradle to roll in or out of the magnet. Seven racks
of services associated with the cradle are mounted on a rolling
platform. As the cradle is rolled out, this platform is rolled
downstream. These services include the HV supplies,
postamp/discriminators, +4\ V preamp supplies, and the gas system.

\begin{figure}
\begin{center}
\includegraphics*[width=140mm]{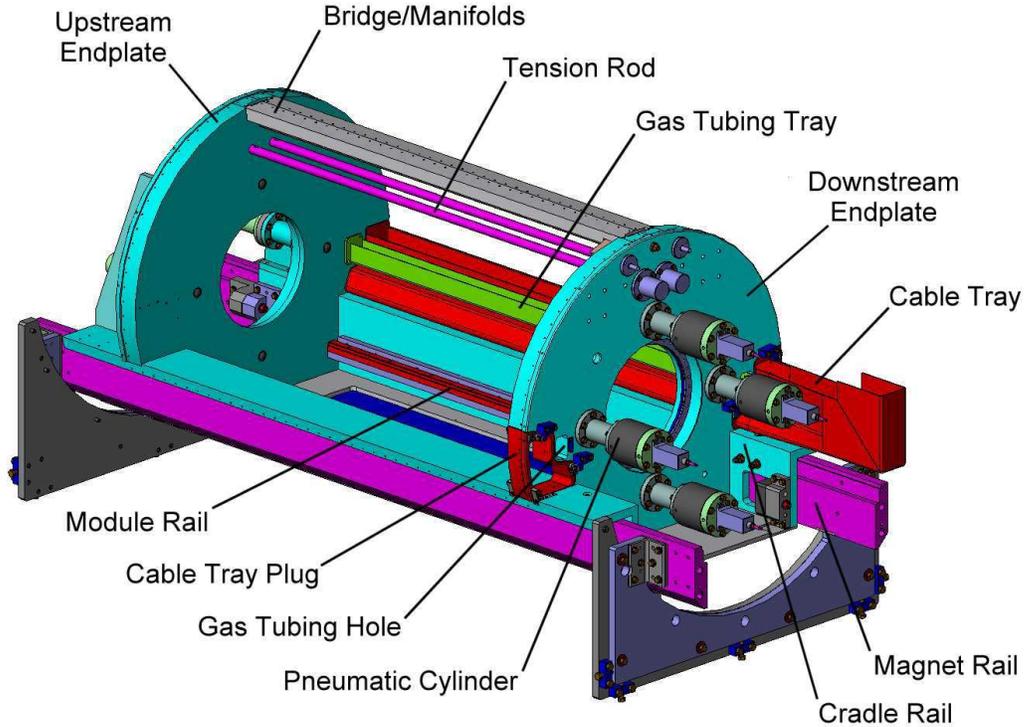}
\end{center}
\caption{
Cradle and magnet rail structure for the TWIST 
modules. One cable tray and gas tray are removed.
}
\end{figure}

The main components of the cradle are the two longitudinal beams,
the two endplates, the ``bridge'' and the tension rods. The beams
have an L shape, with the vertical part being 23 \cm\ high and
the horizontal part 18 \cm\ wide, and both arms are 2.5 \cm\
thick. On the inside face of each beam, there is a module support
rail. As Fig. 4 showed, one of these module rails is V shaped,
the other is flat. The insulating feet on the modules (FR4) match
these profiles and position each module in the cradle in the
($x,y$) plane (or ($u,v$)). The cradle beams continue 27 \cm\
beyond the endplates. Four sets of Be/Cu roller bearings, under
the ends of the beams, allow the cradle to roll on the magnet
rails during installation or removal. During the final few inches
of installation, four 2.54 \cm\ diameter titanium dowels (rounded
tips) on the cradle begin to engage dowel holes and horizontal
slots in aluminum blocks mounted on the magnet rails. With a
dowel hole and precision slot at each end of the cradle, it can
be reinstalled with an ($x,y$) reproducibility of $\leq 0.1\
\mm$, which was as good as the measurement technique used. These
dowels and blocks lift the cradle $\sim\!0.3\ \mm$, so there is
no mechanical conflict with the rollers and they are not under
continuous load. The $z$ location of the cradle is ensured by
having two mating surfaces at the downstream end.

The two cradle endplates are 76\ \mm\ thick aluminum (6061 T6). Each
has a central cutout and step for a window assembly (O-ring
sealed), the windows being 25\ \mum\ doubly aluminized Mylar and
positioned 2.8 \cm\ from the inside endplate face. The endplates
are doweled, bolted and glued to the cradle rails. As discussed
in Section 3, the accurate positioning of the wire planes in the
$z$ direction relies on compressing the four columns of Sitalls
coming from the nineteen modules. There are 113 Sitalls in each
column and they are compressed towards the upstream direction by
four custom pneumatic cylinders mounted on the downstream
endplate. These aluminum cylinders each have a cross-sectional
area of 42.7 \cmsq\ and a maximum operating pressure of 100 psi. We
operate them at 50 psi, where they each compress the Sitall
columns against the upstream endplate with a force of 1,470
N. The gas used in the cylinders is helium rather than air, so
that any small gas leaks don't contaminate the cradle gas.

To further stiffen the upstream endplate against this 5,880 N
load, there are two $51\ \cm \times 25\ \cm \times 2\ \cm$ struts
between this endplate and the L beams. Since the cradle beams are
at the bottom of the cradle, there would still be a tendency for
the endplates to separate at their top edges. To prevent this,
the cradle also has two 2.5 \cm\ diameter aluminum tension rods at
the top. These rods are removed when modules are added or taken
from the cradle. The tension rod design ensures the lengths are reproducible,
so the upstream end plate's position and shape don't change.

As can be seen in Fig. 9, the downstream endplate has a pair of
openings on each side. One is a simple cutout (approximately 34
\cmsq) just outside the cylinders at $x = \pm 260\ \mm$ ($y=0$). They
are for two permanent trays that hold a total of sixty 0.25''
polyethylene gas inlet lines, 56 for the detector layers plus four
spares. These trays and gas lines are permanently installed and
sealed with silicone glue as they pass through the downstream
endplate. If the cradle is removed from the area, these gas lines
are easily disconnected from a nearby external panel and the
short hoses taken with the cradle. The bottom of the cradle has a
12.7\ \mm\ aluminum plate bolted and glued in place. This plate
has a $124\ \cm \times 48 \cm$ cutout and a set of blind tapped
holes for a removable O-ring sealed bottom cover plate. This 6.4\
\mm\ bottom cover plate gives access to the bottom of the cradle
(when it's out of the magnet) to connect or disconnect input gas
lines from the modules.

The other pair of openings on the downstream endplate have much
larger cutouts (approximately 175 \cmsq\ each) and are designed for
cable trays containing the module output mini-coax cables, the
preamp LV cables, the HV cables, module gas and cradle gas
temperature probe cables, and two NMR cables used for field
probes. If the cradle was to be removed from the area, we decided
we wanted to open the cradle and unplug cables from the preamps,
etc., then leave these two cable trays behind. The alternative was
to disconnect all these many cables from the readout electronics,
HV supplies, etc., then coil these long cables back to the cradle
and remove them with it. In particular, the $\sim\!5,000$ signal
mini-coax cables (plus spares) were deemed too fragile for this
process.

In order to leave these cable trays behind, two removable gas
seals were required at the downstream endplate. Our solution was
to make a curved and tapered 76\ \mm\ thick aluminum plug that
surrounded a short section of the cable tray and could mate with
an appropriately shaped cutout in the downstream endplate. A 3.2\
\mm\ thick Poron rubber gasket, between the plug and cutout, is
compressed to $\sim\!2.0\ \mm$ during its installation forming a
good gas seal.

The tray was screwed and glued to the plug, then the cables were
sealed within the tray with silicone rubber. The latter was
difficult with so many cables; air leaks around and through these
cables dominate the cradle leakage. The cradle air leak rate of
$\sim\!2.5$ cc/min is quite acceptable and only results in the
cradle gas containing $\sim\!0.03\%$ air. The cradle has been
removed and installed several times. The removable cable tray
design has shown itself to be simple and reliable.

The ``bridge'' is a removable aluminum structure that is positioned
at the top of the cradle above the modules and tension rods. It
holds the two outlet gas manifolds (see Fig. 4 and Section
9). The sixteen DC modules are connected to one manifold, the
target and two PC modules to the other. The connections are via
soft neoprene rubber bellows, so little or no force is applied to
the modules. The last and major gas seal of the cradle is made by
two 2.4\ \mm\ thick aluminum shells that have gasket seals (Poron)
to the bridge, the endplates and the outside edge of the cradle
L-beams. Each of the shells cover an angle range of approximately
103 degrees on each side of the cradle. 

\subsection{The 100 mm thick upstream FR4 annulus}

Unfortunately, one cannot simply push the most upstream module
against the cradle upstream endplate. The endplate is aluminum
and the modules are glass and FR4. They have thermal expansion
coefficients of \scinot{2.7}{-5}, $\sim\!\scinot{0.5}{-5}$ and
$\sim \scinot{1}{-5}\ (dL/L)/^\circ C$ respectively. The Sitalls
are at $x = \pm 260\ \mm$ ($y=0$) and $y = \pm 260\ \mm$ ($x=0$). So,
over 520\ \mm\ the differential thermal expansion is $>11\
\mm/^\circ C$. For a $10^{\circ}C$ change, the massive 76\ \mm\
thick endplate would be trying to stretch the module $>110\
\mum$. This was considered too dangerous. In addition, pushing
the brittle Sitalls against the aluminum plate was deemed
unwise. As a final concern, the endplate could not be guaranteed
flat enough or stiff enough.

Our solution was to use a 100\ \mm\ thick annulus of FR4 upstream
of the first module. This FR4 annulus has OD of 680\ \mm\ and an
ID 340\ \mm, with four holes milled through it, at $x = \pm 260\ \mm$
($y=0$) and $y = \pm 260\ \mm$ ($x=0$). These four holes each have a
larger diameter step in the upstream face. Four 40\ \mm\ long Sitalls
are glued into the downstream face, using the same optically flat
assembly table used for the detector layers. The holes in the
Sitalls are sealed with 1.6\ \mm\ thick FR4 discs and the holes in
the FR4 annulus are filled with a mixture of epoxy and fine
sand. Finally, 19\ \mm\ thick hard brass inserts (naval brass)
are glued and screwed to the upstream face of the annulus.

This system creates a coplanar set of Sitalls for the upstream
module to be pushed against, and transfers the four 1,470 N loads
to the annulus. Because the FR4 has a thermal coefficient close
to glass, the module should not be stressed. Being 100\ \mm\
thick, it is very rigid. The sand-filled epoxy creates a smooth
load transition to the FR4 and has less compression than regular
epoxy. Contact to the brass inserts is made with four 25\ \mm\
diameter rounded aluminum bumpers that are mounted on the
outside of the upstream endplate and protrude into the
cradle. Their $z$ position is adjustable and lockable. The
threaded region is 22 \cm\ from the inside surface of the
cradle. This distance allows the bumpers to flex slightly in the
($x,y$) plane, to minimize the effects of the different thermal
coefficient of FR4 and the aluminum endplate. The FR4 annulus has
suitable feet machined into it, for the V shaped and flat rails
inside the cradle.

\subsection{The downstream FR4 piece}

At the downstream end of the cradle, the situation is quite
different. It is still important to provide a transition for the
four 1,470 N loads from the pneumatic cylinders to the brittle
Sitalls. The issue of differential thermal coefficients is also
the same as for the upstream end of the cradle. However, while
great stiffness of the upstream annulus was an asset, at the
downstream end of the cradle, compliance is desired. The reason
is obvious; we want the four 1,470 N loads transferred to the
downstream module's Sitalls. A stiff annulus, if its four Sitalls
were not perfectly coplanar, might not transfer those four loads
evenly.

Our solution is to have four 80\ \mm\ diameter, 100\ \mm\ thick
FR4 rods mounted on a 6.4\ \mm\ thick FR4 annulus. Each of these
FR4 rods has a through hole, with a 40\ \mm\ long Sitall, and brass
insert glued into it, just like those of the upstream
annulus. These four 80 mm diameter FR4 rods provide the force transitions; the
6.4\ \mm\ thick FR4 annulus supports them, but is quite
flexible. The pneumatic cylinders, mounted on the downstream side
of the endplate, each have a 25\ \mm\ diameter rounded aluminum
bumper attached to its piston.

These bumpers protrude into the cradle. This distance from
the brass insert to the piston allows the bumpers to flex
slightly in the ($x,y$) plane, minimizing the effects of the
different thermal coefficient of FR4 and the aluminum
endplate. The 6.4 mm thick FR4 annulus has suitable feet machined into it, for
the V-shaped and flat rails inside the cradle.

\subsection{Compression of the 19 modules}

The stack of 19 modules is compressed by the four pneumatic
cylinders. At their maximum pressure of 100 PSI, each of these
cylinders pushes its Sitall column with a force of 2,940
N. Compression tests were performed  to check that we understand
what this compression of the modules is actually doing.

With such tests in mind, the pneumatic cylinders were designed
with the shaft protruding from both ends of the cylinder. In this
way, dial gauges could be mounted outside the cradle to measure
the movement of the cylinders. Unfortunately, since the gauges
have to be mounted on the cylinder casing, their measurements
also include movement of the cradle endplates (especially the
downstream endplate). This cradle movement is about
eight times larger than the compression of the Sitalls (see top
curve in upper part of Fig. 10). To subtract the effect of the
movement of the cradle, we substituted four aluminum pipes for
the detector stack. The response was approximately linear,
especially for $f >1,100 N$. This data was corrected for the
expected compression of the pipes (\scinot{1.46}{-3}\ \mum/N) and
used as the deflection characteristic of the cradle (see bottom
curve in upper part of Fig. 10). The lower curve in Fig. 10 shows
the compression tests of the module stack, after subtraction of
the cradle deflection.

\begin{figure}
\begin{center}
\includegraphics*[width=77mm]{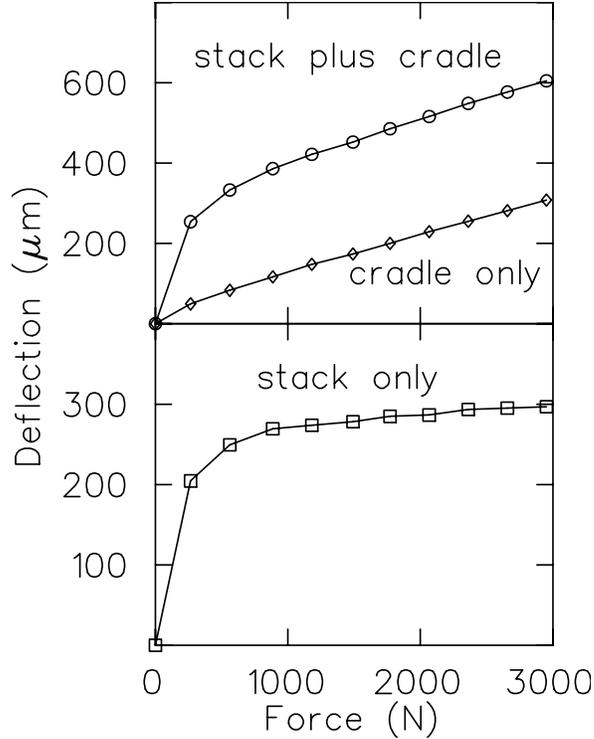}
\end{center}
\caption{
Compression of module stack by pneumatic cylinders. The upper
part shows plots of responses for the cradle as well as for the
module stack plus cradle. The lower plot shows the derived stack
response.}
\end{figure}

As can be seen, the stack of 19 modules shows an initial large
compression, which becomes linear for $f > 1,400$ N. This response is
understandable; the modules have to be pushed into contact and
then the increased force hopefully drives them into nearly
optical contact of their flat surfaces. After optical contact,
the expected slope of the data should be determined by the
elastic modulus of the Sitall ($\sim\!\scinot{5.7}{10}$ Pa), the total inside
length of the cradle (1,483\ \mm) and the area of the Sitalls
(11.6 \cmsq). Assuming the 171\ \mm\ of non-Sitall FR4 end pieces act
like the 1,324\ \mm\ of Sitall, one gets a calculated slope of
\scinot{2.24}{-2}\ \mum/N. The slope of the measured compression in
Fig. 10 is even less, only \scinot{(1.35 \pm 0.12)}{-2}\ \mum/N. This test
should only be taken to indicate that our operating force of
1,470 N on each Sitall column is reasonable to properly compress the
Sitalls in the nineteen modules.

A second study gave us a direct measurement of the length of
the detector stack. As there is no access to the Sitall faces when the
modules are compressed, four brass annuli were installed on the
40\ \mm\ long Sitalls of the dense stack base plates (visible on
Fig. 3). Since these Sitalls extend $\sim\!30\ \mm$ beyond the FR4
gas box, there was room for the 10\ \mm\ thick annuli. They were
positioned $2.000 \pm 0.005\ \mm$ from the ends of the Sitalls
using precision spacers.

These surfaces were then accessible on the upstream and downstream
dense stack module, allowing a direct measurement of most of the
length of the detector stack, including the target module and all
sixteen of the drift chamber modules.

The distance between these upstream and downstream brass surfaces
was measured using a dial gauge mounted in a long aluminum tube
holder. This one meter long ``aluminum dial gauge'' was obviously
subject to temperature changes and needed calibrating. This was
done by repeatedly measuring the length of a calibrated custom
gauge block. This gauge block was $1080.030 \pm 0.005\ \mm$ long
and made of Invar, which has a low thermal coefficient of
$\scinot{1.5}{-6}\ (dL/L)/^{\circ}C$.

Measurements were made at 1,470 N force, then the two measured
$\sim\!2\ \mm$ brass-Sitall distances were added. This gives the
total length of the central 17 modules (end PCs not
included). The value was $1083.785 \pm 0.025\ \mm$. The
calculated lengths of these columns, using the measured
thicknesses of the Sitalls from our data base, is $1083.786 \pm
0.002\ \mm$. This value should be reduced by the calculated
compression of the Sitall column at 1,470 N, which is 24\
\mum. The final value is $1083.762 \pm 0.005\ \mm$. The
difference between this and the measured value is 23\ \mum,
within the $\pm 25\ \mum$ uncertainty.

This excellent agreement confirms that the columns of Sitall are
properly compressed and we know the length of the detector
assembly with a precision considerably better than 50\ \mum. With
the ``aluminum dial gauge'' in place, the cylinder forces were
increased from 1,470 N to 2,940 N and the reduction in the stack
length recorded. The measurement of this change is more accurate;
the value was $27 \pm 5\ \mum$. This is in a good agreement with
calculated value of $24\ \mum$.

Knowledge of the detector stack length within 50\ \mum\ means
that relative position of each 4\ \mm\ Sitall in the stack is
known with a precision of a few microns, an accuracy much higher
than is required.

\section{Gas system}
\label{GasSystem}

Helium/nitrogen ($\sim\!97\!:\!03$) flows through the cradle and
between the modules, and the first and last cathode foils in each
module act as the module gas windows. Two gas systems are
required for the TWIST modules, one for the DME gas of the
sixteen DC modules, the second for the CF$_4$/isobutane gas for
the target and the two PC modules (T+PC). These gas systems are required
to provide stable gas flows to each of the 56 individual
detectors while maintaining a very low differential pressure
between the modules and the helium/nitrogen gas of the cradle.

\subsection{Description}

Flow control for the DC system is provided by a pressure
regulated input manifold feeding DME through 44 precision needle
valves to the 44 individual detector layers. Each of the 44 input
flows is continuously monitored by inexpensive mass
flowmeters. The pressure between the common DC output manifold
and the cradle gas is measured with a precision differential
pressure transducer. This transducer signal is used to adjust the
output flow to the DME system output pump, thus realizing
differential pressure control. To ensure all sixteen DC modules
have a common differential pressure with respect to the cradle,
the short neoprene bellows ($\sim\!0.8\ \cmsq \times 7\ \cm$)
connecting the outputs to the manifold, and the output manifold
itself ($\sim\!10\ \cmsq \times 140\ \cm$), have very low flow
impedance. Solenoid valves prededing the DME input manifold and
following the outlet manifold allow the 16 DME modules to be
isolated, to provide protection against accidental over or under
pressure. These valves are programmed to close if the
differential pressure exceeds $\pm 150$ mTorr from the desired
setpoint pressure. A pressure relief bubbler is also set to vent
at $\sim\!500$ mTorr with respect to atmosphere, providing a
final ``fail safe'' protection.

The flow and pressure control features of the (T+PC) gas system
are identical to the DC gas system described above. The only
differences are that the gas is CF$_4$/isobutane (80:20), there
are only 12 detector layers (4 in the target and 4 in each of the
two PC modules), and a lower quality differential pressure
transducer is used. Because CF$_4$ is an expensive gas, 80\% of
the chamber output flow is filtered, mixed with the incoming
fresh supply, and recycled.

The helium/nitrogen ($97\!:\!03$) gas is supplied from a pressure
regulated source and uses a simple mechanical flowmeter with a
manual needle valve. The flow rate is typically 1 l/min to the
cradle. The helium/nitrogen from the cradle is exhausted through
a 1'' diameter copper tube to vent at an elevation 99 \cm\ below
the midplane of the modules. The low impedance output tube is
necessary to reduce cross coupling of pressure fluctuations
between the (T+PC) and DC pressure control systems
that can result from pressure fluctuations being transmitted
through the windows of the modules of one system, through the
cradle gas and into the windows of the modules of the other
system. Venting the cradle gas to atmosphere 99 \cm\ below the
midplane creates a static overpressure of $\sim\!80$ mTorr with
respect to atmosphere at the modules' midplane. A pressure relief
bubbler, identical to those used in the (T+PC) and DC
gas systems, provides protection against accidental over
pressuring of the cradle.

\subsection{Pressure control}

A significant challenge for the gas system design was to maintain
the external cathode foils as flat as possible and with a
positional stability of $\pm 50\ \mum$ or less. There are three
sources of differential pressure that can deflect the module
windows:
\begin{enumerate}
\item Electrostatic pressure due to the anode-to-cathode
electrostatic attraction.
\item Gravitational pressure due to the vertical 
orientation of the cathodes and the difference in 
density between gasses on either side of the first 
and last cathodes.
\item Gas pressure applied by the gas supply 
systems.
\end{enumerate}
To minimize the impact of all three effects, it was necessary to
make the foil tension as high as was practical. Preliminary
tension tests of stretched 339\ \mm\ diameter, 6.3\ \mum\ thick
foils (doubly aluminized) revealed that after being stretched and
glued to the cathode frame, the foil tension relaxes
exponentially with a time constant of $\sim\!100$ days, to an
asymptotic value $\sim\!75\%$ of the initial tension. Figure 11
shows the relaxation of a test foil.

\begin{figure}
\begin{center}
\includegraphics*[width=77mm]{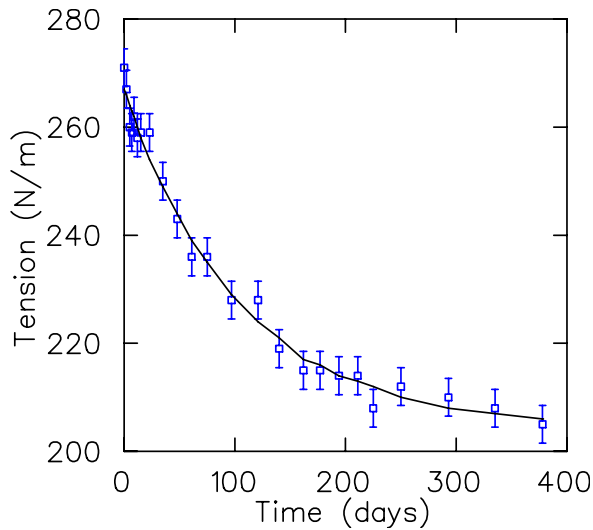} \end{center}
\caption{
Exponential relaxation of tension after 6.3\ \mum\ aluminized
Mylar cathode foil is stretched and glued. Foil tension relaxes
to 206 N/m (77\% of its original value) with an exponential time
constant of 102 days.
}
\end{figure}

The first 59 production cathode foils were stretched to an
initial tension of $\sim\!340$ N/m. Due to concerns of foils ripping
from the high tension, the final 107 cathode foils were stretched
to a lower initial tension of $\sim\!255$ N/m. After gluing to the
cathode frame, the tension in each foil was measured and the
expected final relaxed tension was calculated. From these
measurements, the average expected final tension of all the foils
was calculated to be $206 \pm 20$ N/m. 

A tension of 206 N/m corresponds to an average foil deflection
rate of 4.7\ \mum/mTorr, at the center of the circular foil. This
implies that to keep the position of the centre of the foils
stable to $\pm 50\ \mum$, we would need to keep the differential
pressure across the window foils constant to $\pm 11$ mTorr.

After assembly, the pneumatic capacity of each module was
measured by flowing gas into the closed module and monitoring the
resulting rate of change of pressure. Pneumatic capacity ($C$) is
defined by $F = C dP/dt$, where $F$ is the volume flow rate at
STP and $dP/dt$ the rate of change of pressure in the closed
volume. From these measurements, we calculated the average
tension of the window foils of each module and the average
deflection rate of these foils ($dz/dP$). The average measured
capacity of the 19 TWIST modules was $0.40 \pm 0.02$ cc/mTorr
corresponding to an average tension in the foils of $221 \pm 11$
N/m, and a deflection rate of $4.5 \pm 0.2\ \mum/$mTorr.

Since roughly one year ($\sim\!3.5$ tension-decay time constants)
passed between the initial foil tension tests and their final
assembly into modules, the results of the initial foil tension
tests and the module tests are consistent.

The total capacity of the 16 DC modules (in the cradle) was also
measured several times during various running periods of the
TWIST spectrometer. A repeatable pattern emerged. After one day
of exposure to DME the total capacity would be $\sim\!11$
cc/mTorr. The total capacity would continue to increase over the
next few weeks until it stabilized at $\sim\!16.5$ cc/mTorr,
significantly higher than the expected $16 \times 0.4 = 6.4$
cc/mTorr. After exposure to air or argon for a few weeks, the
total capacity would again decrease to $\sim\!11$ cc/mTorr. The
DC operating total capacity of 16.5 cc/mTorr implies an average
foil deflection rate of 11.6\ \mum/mTorr, which requires a
differential pressure stability of $\pm 4.3$ mTorr to maintain foil
position stability of $\pm 50\ \mum$.

The total capacity of the target and two PC modules was also
measured during the various running periods. Total capacity
appeared to be stable at $\sim\!1.45$ cc/mTorr regardless of time
of exposure to CF$_4$/isobutane, argon/isobutane, or air. This is
slightly higher than the expected total capacity of $3 \times 0.4
= 1.2$ cc/mTorr, and implies an average foil deflection rate of
$\sim\!5.4\ \mum/$mTorr.

From these results, in particular the responses to the different
gas mixtures, we suspect that the DME is being absorbed by the
Mylar foils of the DC modules, causing the foil tension to
relax. Some of this relaxation appears to be permanent. Bagaturia
et al. \cite{Bagaturia} have also noticed foil relaxtion
associated with DME. In their case, the foils were Kapton GEM
detectors.

A high precision pressure transducer is used to measure the
differential pressure between the DC modules and the cradle
gas. At our typical operating differential pressure the
manufacturer's specifications imply a temperature drift of 0.023
mTorr/$^{\circ}C$. This is sufficient for our requirements, where
typical extremes of temperature are less than $\pm 6\
^{\circ}C$. The foil stability requirements for the target and
PCs are considerably less strenuous and consequently a less
precise (less expensive) pressure transducer is used on the
(T+PC) system. The manufacturer's specifications indicate a
temperature drift of
1.08 mTorr/$^{\circ}C$ for this gauge at our typical operating
differential pressure.

For each gas system, the output signal from the pressure
transducer is connected to a PID controller that in turn controls
the output mass flow controller (MFC) preceding the exhaust pump,
thereby controlling the differential pressure between the modules
and the cradle volume. As shown in Fig. 12, the control stability
obtained is the order of $\pm 0.5$ mTorr. Thus, the combined
transducer temperature drift and control instability contribute
an error less than $\pm 1.0$ mTorr for the DC system,
corresponding to only $\pm 12\ \mum$ foil
stability. Unfortunately, these are not the only sources of
differential pressure instability.

\begin{figure}
\begin{center}
\includegraphics*[width=77mm]{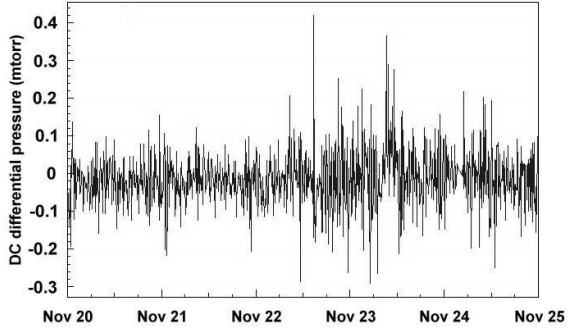}
\end{center}
\caption{
Typical pressure control stability over a five day period.
}
\end{figure}

Space constraints and high magnetic fields around the cradle
required that the pressure transducers be located several meters
from the measurement points. ``Blind'' 0.25'' OD copper tubes
that traverse rising, horizontal and falling sections, connect
the transducers to their measurement points. These three pressure
sensing tubes are initially flushed out with the appropriate gas
mixtures (DME, CF$_4$/isobutane, or helium/nitrogen) and then
left closed at the pressure transducer end. Accurate knowledge of
the differential pressure at the centre of the chambers depends
on accurate knowledge of the gas density in these tubes. A 1\%
change in density due to atmospheric pressure or ambient
temperature change causes a 1.1 mTorr change in the measured DC
differential pressure. Since atmospheric pressure and the gas
temperature at several locations in the pressure sensing tubes
are monitored, these changes can be compensated for.

More problematic are changes in gas composition due to diffusion
or small leaks into the pressure sensing tubes. During the 2002
running period a technique was developed for periodically
measuring the gas density in the pressure sensing tubes. These
measurements revealed changes of up to 40 mTorr in the actual
differential pressure at the centre of the window foils of the DC
modules. Subsequent intensive leak checking revealed some small
leaks in the pressure sensing tubes. These leaks were fixed and
during the 2003 running period the actual DC differential
pressure was stable to $\pm 7$ mTorr implying a foil position
stability of $\pm 80\ \mum$.

To further improve the positional stability of the window foils,
we are considering a modification, so that the three pressure
sensing tubes are continuously flushed with small flows of the
appropriate gas mixtures. These small flows would start at the
pressure sensors, pass through the sensing lines and into the
chamber gas manifolds and the cradle.  Flows of $\sim\!1$ cc/min
would not cause significant flow related pressure drops in the
pressure sensing tubes.

Currently we are calculating the foil deflections based on
pressure, temperature and chamber capacity measurements. An
independent indication of the foil deflections is desirable. We
are investigating the possibility of using online TDC data from
the detectors to independently monitor external cathode foil
deflections.

\subsection{Gas composition stability}

Since helium/nitrogen cradle gas surrounds the wire chambers,
diffusion of helium into the chambers is a major source of
contamination.  GARFIELD simulations predict that a 1\% change in
helium concentration in the DME (from say 1\% to 2\%) would cause
roughly a 1\% change in electron drift times. Preliminary tests
of numerous pieces of 6.3\ \mum\ aluminized Mylar foil determined
an average helium diffusion rate of \scinot{(3.6 \pm 1)}{-6}
cc/(s\ Torr\ m$^2$) or about 0.015 cc/min through each window
foil. Our normal flow rate of $20 \pm 2$ cc/min per detector layer
results in a helium concentration of 0.075\% with a stability of
$\pm 0.008\%$ for the (UV) modules. Since 80\% of the (T+PC) gas is
recycled, the helium contamination in the CF$_4$/isobutane due to
diffusion through the window foils is expected to stabilize at
$(0.19 \pm 0.02)\%$.

Small leaks through pinholes in the cathode foils or holes in the
module gas boxes could easily exceed diffusion through the window
foils. For this reason all modules were leak checked after final
assembly by filling them with helium to $\sim\!+500$ mTorr with
respect to atmosphere. The inlet valve was then closed and the
differential pressure to atmosphere was monitored for at least 15
minutes. Leak rates were calculated from the module's measured
capacity and any resulting changes in differential
pressure. Continuous measurements of gas temperature and
atmospheric pressure allowed us to compensate for these
effects. Only (UV), target and PC modules with measured leak rates
less than 0.3 cc/min were installed in the cradle. Due to the
four times higher operational total flow through dense stack (DS)
modules, leak rates up to 0.6 cc/min were accepted for the DS
modules. The average measured helium leak rate of the 14 (UV)
modules was $0.14 \pm 0.07$ cc/min per module. This corresponds
to an average helium concentration of 0.35\% with a concentration
instability of $\pm 0.04\%$ at our nominal flow rate and flow
rate stability. This instability should cause a similar
uncertainty of $\pm 0.04\%$ in the drift velocity, which is well
within our required tolerances.

During running periods we periodically estimated the total leak
rates of the DME, CF$_4$/isobutane and cradle volumes. This was
done by increasing the differential pressure in the volume to be
measured by a few hundred mTorr, isolating the volume and
monitoring the pressure over a period of time. These tests are
sensitive to leak rates in the modules and all their connecting
tubing between the isolation valves at the gas racks and the
volumes under test. The measurements were primarily used to
detect changes in leak rates following reinstallation of
modules. The typical measured total leak rate for the DCs (DME)
was 4.5 cc/min or 0.28 cc/min per module. The cradle leak rate
typically measured approximately 60 cc/min. The differential
pressure in the CF$_4$/isobutane volume typically showed an
increasing pressure when isolated. We suspect this effect is due
to diffusion of helium or air into the detector end of the
(T+PC) pressure sense line.

The CF$_4$/isobutane (80:20) for the (T+PC) system is mixed with
MFCs having a manufacturer's specified accuracy of $\pm 1\%$ of
full scale. At the mixing flow rates employed, this results in a
mixture of $(80 \pm 0.8)\%$ CF$_4$ and $(20 \pm 0.2)\%$
isobutane. Since pure DME is used for the DCs mixture accuracy is
not an issue. The manufacturer's specifications for DME and
isobutane supplies are 99.5\% purity with typical impurities of
n-butane and other alkanes. The CF$_4$ used is 99.95\%
pure. Oxygen contamination of all gas supplies is measured before
use and typically found to be less than 20 ppm. The oxygen
concentration in the exhaust gasses was measured at various times
during running periods. Typical oxygen concentrations were 75 ppm
in the DCs, 50 ppm in the (T+PC) and 500 ppm in the
cradle exhaust gas. For a 1 l/min flow through the cradle, this
500 ppm oxygen concentration implies an oxygen flow in of 0.5
cc/min, and an air leak rate of 2.5 cc/min. To compensate for
this implied leak, the nitrogen content of the cradle input gas
was reduced from 3\% to 2.75\%, so the cradle gas is actually
helium/nitrogen/air ($97\!:\!2.75\!:\!0.25$).

\section{Readout electronics}
\label{ReadoutElectronics}

On all modules the cathode foils are grounded and positive high
voltage is applied to the wires. This allows each detector layer
to have its own operating voltage or to be turned off. The
signals are brought through the gas box wall at high voltage and
then decoupled on the preamplifier boards. Two service PCBs are
mounted on each module gas box (in the readout arcs). They
provide mounting and +4 V distribution for the preamplifiers.

All signal, preamplifier power, and HV cables are permanently
fixed in two cable trays and sealed with a silicone rubber. The
cable trays can be disconnected from the cradle and chamber stack
when the detectors need to be moved out of the experimental area
for service or tests (as discussed in Section 8).

We use a preamplifier developed at Fermilab for use at their
Colliding Detector Facility \cite{Yarema}. This preamplifier is used for
all the TWIST chambers and has a gain of 1 mV/fC and a dynamic
range of -400 fC to +20 fC. Both 16 and 24 channel versions of
this preamplifier are used on the detector modules.

All signals from the DCs, target and PCs go to
post-amplifier/discriminators via 9.5 m long micro-coaxial
cables. These custom made post-amplifier/dis\-crim\-inator modules
have sixteen channels in a single width CAMAC unit and are housed
in twelve CAMAC crates within two racks on the service platform
next to the spectrometer. Each CAMAC crate contains up to 24 of
these units plus a custom made controller module. This controller
module interfaces with the TWIST Slow Control system and allows
the adjustment of discriminator thresholds and application of
test pulses. Temperature and power supply voltages are also
monitored via this controller module.

The discriminator circuit produces a differential ECL logic, time
over threshold output. Our operating threshold for the DC postamps
is typically 150 mV. With their gain of twenty, this means a
threshold equivalent to 7.5 mV at the preamp output. The VTX
preamplifier has a gain of 1 mV/fC and our DC gas gain is
estimated at \scinot{(1.8 \pm 0.2)}{4} (at 1950 V). This implies
the effective threshold is $\sim\!2.5$ electrons collected from a
passing track. A more direct measurement indicated that detection
of single electrons produced $\sim\!4.5$ mV pulses from the
preamps, so our threshold was in fact $\sim\!1.6$ electrons. The
latter is more likely, since the specified preamp gain will
depend on frequency. Under these operating conditions, the noise
rate was measured and found to be negligible ($\leq 10$ Hz/wire).

These discriminator ECL logic signals are sent to the FASTBUS
TDCs (LeCroy model 1877) via 15 m long, 16-pair flat-twisted
cables. The cables are bundled and wrapped with copper laminated
Mylar foil to reduce oscillations caused by RF radiation. The
TDCs are multihit type and have 0.5 ns resolution. They are
operated in Common Stop mode.

The FASTBUS TDCs are housed in 2 crates each containing an
SIS4100NGF FASTBUS to VME interface. Each interface contains a
Motorola MVME 2306 PowerPC which is responsible for transferring
the TDC data to the data acquisition system \cite{Poutissou}
through an Ethernet connection. The FASTBUS crates are positioned
8 m from the service platform, to eliminate the interference from
their switching power supplies.

\section{Efficiency}
\label{Efficiency}

The first efficiency tests on the drift chambers were conducted
with the spectrometer magnet off, using a 120 MeV/c pion beam.

The efficiency code uses tracking information to determine which
wire in a given plane is expected to display a hit. Once a track
is successfully reconstructed, the track parameters are used by
the efficiency algorithm to traverse through the detector stack and
find the entrance and exit points of the track through each
detector layer. This information is then converted into cell
numbers, and the plane is searched to determine whether the TDCs
corresponding to these cells (or the neighboring cells) recorded
a hit. In this context, the term efficiency refers to the
intrinsic efficiency of the chamber. This efficiency depends on
the gas properties, cell geometry and construction (cell size,
wire thickness, presence of drift wires, etc.), high voltage and
threshold.

To determine an operating point for the DCs, data were
obtained with the high voltage of approximately half the modules at
1900 V while the high voltage on the other half was varied from
1600 V to 2000 V in steps of 50 V. Figure 13 shows the efficiency
as a function of high voltage averaged over all DC planes for which
the HV was being varied. The figure insert is an expanded view of
the ``plateau region'', showing that the efficiency reached a
value greater than 99.95\%.

\begin{figure}
\begin{center}
\includegraphics*[width=77mm]{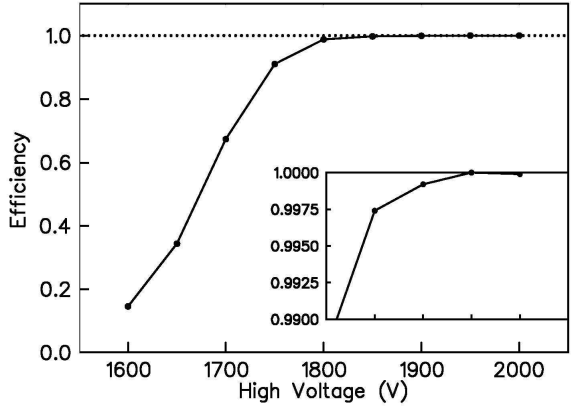}
\end{center}
\caption{
DC efficiency as a function of HV.
}
\end{figure}

In order to check possible plane-to-plane variations, the
efficiency of each plane as a function of high voltage was
calculated. Variations at 1900 volts were found negligible with
all functioning planes showing efficiencies better than
99.8\%. Differences in the shape of the efficiency as a function
of high voltage in the region of interest were also found to be
negligible, thereby allowing a single operating high voltage for
all DC chambers. This is 1950 V. Similarly, the operating voltage
of the target and PC modules was determined to be 2050 V.

Since variations in efficiency across a plane may introduce
variations in energy and angular acceptance, the efficiency
algorithm was also expanded to calculate a wire-by-wire
efficiency. Variations in efficiency from wire to wire were also
found negligible. The chamber efficiencies per plane are
continuously monitored during data taking to ensure stability.

The algorithm used for efficiency calculations was carefully
tested for possible biases. A fraction of the hits was rejected
right after the unpacking of the TDCs, with the rejection factor
varied by different amounts. In particular, to test the
efficiency code for possible small biases, as well as
sensitivities to inefficiencies of the order of \scinot{1}{-3}, the
rejection factor was set to 0.001 and 0.002 and the difference in
the calculated efficiency between the two cases was
computed. This resulted in the expected efficiency difference of
0.001.

Since 120 MeV/c pions deposit more energy than muon-decay
positrons, and since the Lorentz angle may in principle cause the
efficiency to deteriorate when the field is on, the DC efficiency
was also calculated using decay positrons at the operating high
voltage of 1950 V. The results were very similar to the pion data
reported above, thereby showing no deterioration in efficiency.

\section{Alignment}
\label{Alignment}

The chamber construction techniques give a high precision in
inter-plane alignments within a detector layer (i.e., wire
positions, foil positions, etc.), as was discussed in Section
6. The use of Sitall spacers and the cradle compression system
also gives extremely high precision in the $z$ position of the wire
planes (as was discussed in Sections 2, 3, and 8).

However, the module assembly and mounting does not allow such a
high precision in the transverse chamber positions (($x,y$) or
($u,v$)) or rotations around the $z$ axis. A high precision,
however, is not required, since straight tracks allow for a high
precision determination of both.

The TWIST alignment code uses 120 MeV/c pion tracks obtained with
the spectrometer magnet off to determine the transverse plane
alignments and rotations around the $z$ axis. For translational
alignments, only tracks close to the center of the chamber are
selected to make sure that translational and rotational
alignments do not mix at a significant level. Each track is
fitted using a Kalman filter, and the means of the tracking
residuals for each plane (except two for each alignment direction
which are kept fixed) are used to adjust the transverse position
of that plane. This process is repeated until all plane positions
converge.

Figure 14 shows the Monte Carlo convergence of wire plane
differential offsets (difference between Monte Carlo offsets and
means of the tracking residuals), for the upstream 22 DC
planes. The upper part shows the translational differential
offsets of the tracking residuals at the end of each iteration
for Monte Carlo data where translational misalignments of twice
those obtained from data were introduced. The iteration procedure
converges nicely, as is evident from the figure, and the
precision to which the alignment code is able to recover these
misalignments determines the accuracy of this procedure to be
$\leq 10\ \mum$.

\begin{figure}
\begin{center}
\includegraphics*[width=77mm]{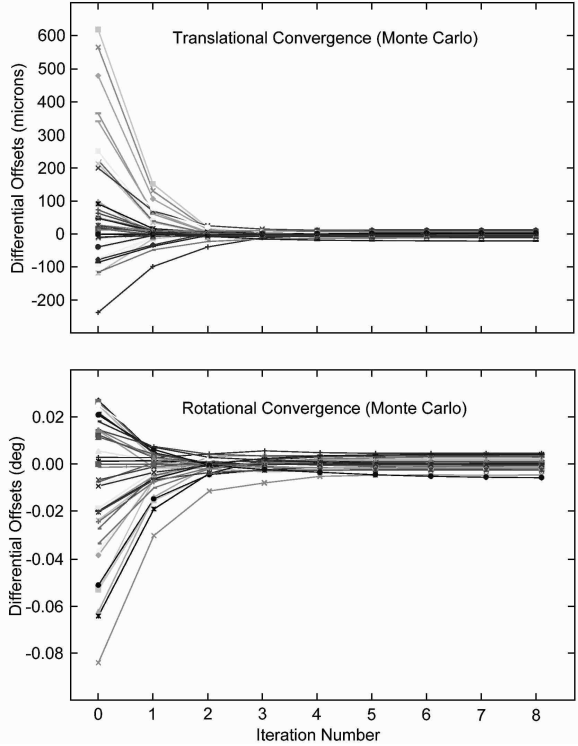}
\end{center}
\caption{
Monte Carlo convergence of translational (upper) and rotational
(lower) wire plane offsets, for the upstream 22 DC planes.
}
\end{figure}

For rotational alignments, the tracking residuals are computed in
bins along the wire length, and the means of the tracking
residuals along the wire length are converted into a rotation
angle. This angle is then used to introduce a plane rotation
correction, and the process is iterated until the plane rotations
converge. The lower part of Fig. 14 shows the rotational
differential offsets at the end of each iteration for Monte Carlo
data where rotational misalignments of twice those obtained from
data were introduced. The figure demonstrates convergence of the
alignment process. The alignment code was also tested using Monte
Carlo data and verified to be independent of the starting plane
positions and independent of the alignment axis, defined by the
fixed planes. The precision is better than 0.02 degrees.

For the actual chambers, this code was used to determine the
corrections for the 22 upstream DC planes and also the 22
downstream DC planes. The corrections were found to quickly
converge. The derived corrections were reproducible to better
than 10\ \mum\ for translations and better than 0.02 degrees for
rotations, independent of the starting values of the
misalignments. When the same 22 upstream DC planes were analyzed
two months later, the new translational corrections differed by
less than 9\ \mum\ for the worst planes ($\sigma \sim 3\
\mum$). This is quite consistent with the technique's accuracy of
$\sim\!10\ \mum$.

The required wire-plane corrections were found to vary up to
$\sim\!300\ \mum$ for translations and up to $\sim\!0.05$ degrees
for rotations. These plane position corrections reflect relative
plane-to-plane alignments, and their magnitude depends on the
planes which were fixed to determine an alignment axis. The
corrections, therefore, do not translate directly into actual
plane positions. While the planes are expected to be positioned
to an accuracy of $\sim\!80\ \mum$ within a module, the
module-to-module misalignments can be up to few times larger.

Since this alignment procedure requires the magnet to be off,
calibrations runs are made at the beginning and end of each
running period to ensure there are no changes. In addition, there
is an optical alignment system on the cradle to monitor its
position. This system uses the end of an optical fiber as a
target. A halogen light illuminates the other end of the fiber
and the light emitted from the target end is focused with lenses
and viewed with a CCD camera element. Four such targets are
mounted on the bottom surface of the cradle, two widely spaced
near each endplate and a fifth target is near beam height and
pointing horizontally at right angles to the beam direction (the
$+x$ direction). These five targets are viewed through holes
drilled in the magnet yoke. The optical system does not touch the
magnet and is firmly bolted to the concrete floor. This system
indicates that the magnet and cradle system is mechanically
stable to $\sim\!50\ \mum$ and not affected by the magnetic field
being turned on or off.

One shortcoming of the alignment procedure using straight tracks
is the inability to determine the direction of the axis of the
detector relative to the direction of the axial magnetic
field. However, this quantity can be determined using positron
helices when the magnetic field is turned on. The hit pattern of
the positrons on each detector plane (i.e., as a function of $z$)
was used to determine this axis to better than 0.035 degrees.

\section{Resolution}
\label{Resolution}

As discussed in previous sections, the high precisions of the wire
plane construction and the Sitall system for $z$ positioning of
the wire planes result in a negligible impact on the overall
spatial resolution of the spectrometer. The transverse plane
positions (perpendicular to the $z$ axis), and rotations around
this axis result in a significant contribution. However, once the
translational and rotational corrections of each wire plane are
applied, their contribution to the spatial resolution of the
spectrometer are small.

There are two major contributions to the spatial resolution of
the chamber; the properties of the drift cell and the mechanical
precision of the detector assembly. The properties of the drift
cell are mainly determined by the choice of drift gas. The high
ionization density, low drift velocity and small Lorentz angle
make DME a desirable choice for this experiment.

To determine the resolution of the DC planes, a subset of the
upstream 22 DC planes were chosen. To define the tracks properly,
but minimize multiple scattering, only 8 planes were used (four U
and four V). The planes chosen were DC\#7 through DC\#14. These
would be used to test the resolution in a ninth plane
(DC\#6), a V plane. DCs \#6 to \#8 are the last three planes in the dense
stack module and the other six are the next three (UV)
modules. This choice of planes means the subset spans only 188\
\mm\ (minimizing the effects of multiple scattering) and the
plane not in the track fit is only 4 mm away from the first
plane, so projection errors are also minimized.

As in Sections 11 and 12, we use 120 MeV/c pion data with the
magnet off. Events near normal incidence ($\theta \leq 5$ degrees)
were chosen so we could map the resolution across the drift
cell. For DC\#7 through DC\#14, events were selected that had drift
distances $>0.5\ \mm$, where the resolutions were better. These
tracks were fitted to a straight line and compared with the drift
times in DC\#6, the plane being studied. The STRs (space-time
relations) started with those derived from GARFIELD, but were
allowed to iterate.

Figure 15 shows the final resolutions as a function of distance
from the wire. The resolution is below 50\ \mum\ for tracks more
than 1\ \mm\ from the wire, but closer than that it gets
progressively worse. This deterioration in resolution closer to
the wire is mainly the result of the ionization statistics.

\begin{figure}
\begin{center}
\includegraphics*[width=77mm]{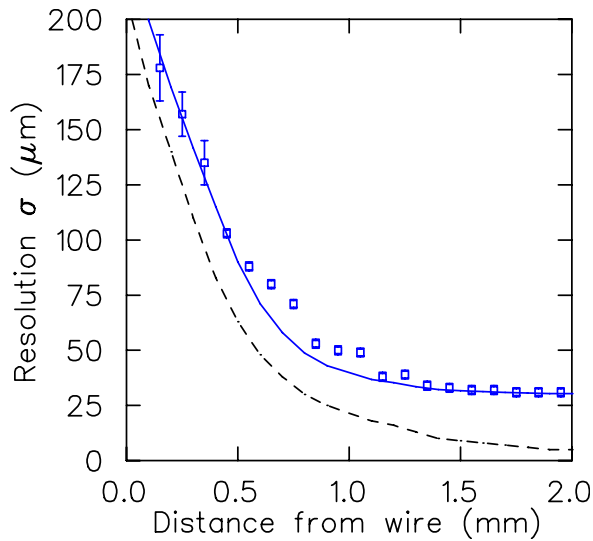}
\end{center}
\caption{
Measured resolution as a function of track distance from the wire
(points with error bars). The dashed curve is a GARFIELD
calculation, while the solid curve also includes the quadratic
addition of 30\ \mum\ spatial resolution plus 1.5 ns time
resolution (see text).
}
\end{figure}

The lower curve shown in Fig. 15 is the resolution computed from
GARFIELD using a threshold of 1.6 electrons, which corresponds to
our best estimate of our actual threshold (see Section 10). While
it shows the resolution deteriorating closer to the wire, it
predicts too good a resolution in most of the cell. This is not
surprising, since this calculation does not include contributions
such as: (a) residual alignment errors ($\leq 10\ \mum$) and (b)
multiple scattering of the 120 MeV/c pions over the 188\ \mm\ distance
of the nine wire planes, and (c) timing jitter associated with
leading edge timing and pulse height variations.

To agree with our data, we would need an added resolution
contribution that was $\sim\!30\ \mum$ at 1.8\ \mm\ and rises to
$\sim\!80\ \mum$ at 0.4\ \mm. For example, a multiple scattering
contribution of $30\ \mum$ and a timing uncertainty of $\sim\!1.5$
ns added in quadrature to the GARFIELD calculation produce good
agreement with our measured resolutions (see upper curve in
Fig. 15). Such contributions, or others, could easily account for
the discrepancy.

These results were obtained with the DC chambers operating at
1900 V. Since then the operating voltages of the DCs have been
raised to 1950 V. Since the resolution is threshold dependent, we
will be collecting more test data with voltages of 1950 V, 2000 V
and 2050 V. At 2050 V, the gas gain should be $\sim\!50\%$ higher.

Cindro et al. \cite{Cindro} used this technique for determining
the resolution of their DME chambers. They concluded that their
threshold was 10 electrons (2.0 clusters of 5 electrons each) and
added a constant value of 20\ \mum\ (not in quadrature) to match
their observed resolution, which were only about $15\ \mum$ worse
than ours.

Figure 16 shows the distribution of tracking residuals for
DC\#6. Since resolution deteriorates for tracks closer to the
sense wire (Fig. 15), this distribution is only for hits having
drift distances greater than 0.5\ \mm. The distribution has FWHM =
$80\ \mum$. The tails extend to $\sim\!250\ \mum$, and undoubtedly have
contributions from; (a) multiple scattering over 188\ \mm\ and 9
detector layers, and (b) hits near the 0.5\ \mm\ rejection distance,
where the resolution is already $\sim\!90\ \mum$.

\begin{figure}
\begin{center}
\includegraphics*[width=77mm]{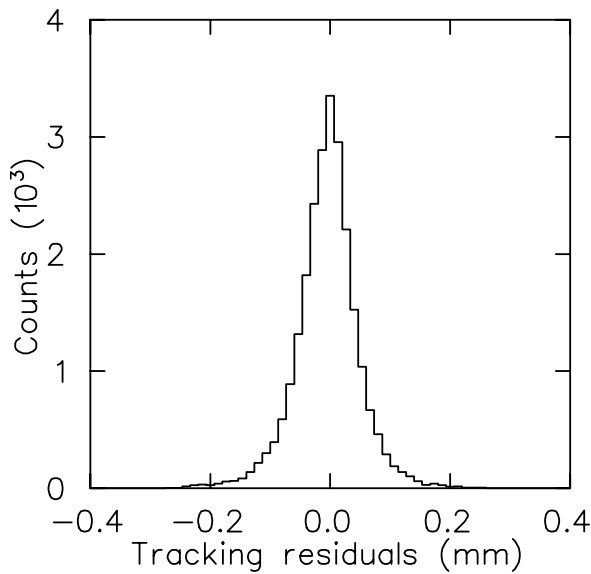}
\end{center}
\caption{
Tracking residual distribution of DC\#6, for drift distance $>
0.5\ \mm$, with FWHM of $80\ \mum$.
}
\end{figure}

The resolution measurements indicate that the TWIST drift
chambers are giving resolutions as good as were expected.

\section{Conclusions}
\label{Conclusions}

The TWIST spectrometer was commissioned two years ago. There have
been many calibration and data collection runs. The operation and the
performance of the TWIST chambers was virtually flawless. There
are no hot or dead wires in the entire spectrometer, containing
5,056 sense wires from 44 DC planes, 8 PC planes and 4 target
planes. Of particular importance, the DC planes operated at full
efficiency ($>99.95\%$).

The detector system has been extremely reliable, with only one
broken wire. This required the PC module to be removed and
replaced with the spare PC module.

Incorporating a set of low thermal expansion Sitall spacers in
every detector layer has resulted in a system where the $z$
positions of each wire plane are known to a few microns and
cumulative tolerances over the 120 \cm\ long tracking region are
less than 50\ \mum. Using glass plate substrates for the detector
layers has also resulted in a cumulative tolerance across the
320\ \mm\ wide active areas of less than $\pm 6\ \mum$.

Within each wire plane the sense wires were strung with excellent
precision ($\sigma =3.3\ \mum$ in the $x$ direction), with very
few wires out of position more than 15\ \mum.

The mechanical system of the cradle was well designed to utilize
the high quality of the module construction and to make
installation and removal of modules (or even the whole cradle) as
straightforward as possible.

\section{Acknowledgements}
\label{Acknowledgements}

We thank the full TWIST collaboration. The design and
construction of the TWIST detectors and the cradle support system
could not have been accomplished without their suggestions,
advice and support. D.R. Gill and N.L. Rodning, as the first
two spokespersons of the TWIST collaboration, deserve special
mention. We thank H.C. Walter and J.A. Macdonald for their
contributions,
G. Stanford for his invaluable engineering advice and A. Prorok
for his excellent design office skills. We thank C.A. Ballard,
M.J. Barnes, S. Chan, B. Evans, M. Goyette, D. Maas,
G.M. Marshall, S. Mullin, J. Soukup, C. Stevens, P. Vincent, and
L. Wampler, who contributed to the construction and operation of
the TWIST chambers and cradle system. We also acknowldge many
contributions by other members of the TRIUMF professional and
technical staff. This work is supported in part by the Natural
Sciences and Engineering Reasearch Council and the National
Reasearch Council of Canada, the Russian Ministry of Science, and
the U.S. Department of Energy.



\end{document}